\documentclass[11pt]{article}
\usepackage{epsfig,amscd,amssymb}
\begin{document}

% ---------------------------------------------------------------------%
%
%
%-------------------------------------------------------------
%
%
\title{Integrable sigma models with $\theta=\pi$}

\author{Paul Fendley\\
Department of Physics\\
University of Virginia\\
Charlottesville, VA 22904-4714\\
{\tt fendley@virginia.edu}
}
\maketitle

\begin{abstract}
A fundamental result relevant to spin chains and two-dimensional
disordered systems is that
the sphere sigma model with instanton coupling $\theta=\pi$
has a non-trivial low-energy fixed point and a gapless spectrum.
This result is extended to two series of sigma models with $\theta=\pi$:
the $SU(N)/SO(N)$ sigma models flow to the
$SU(N)_1$ WZW theory, while the $O(2N)/O(N)\times O(N)$ models flow
to $O(2N)_1$ ($2N$ free Majorana fermions). These models are integrable,
and the exact quasiparticle spectra and $S$ matrices are found.
One interesting feature is that charges fractionalize when $\theta=\pi$.
I compute the energy in a background field, and verify that
the perturbative expansions for $\theta=0$ and $\pi$ are the same
as they must be. I discuss the flows between the two sequences of models,
and also argue that the analogous
sigma models with $Sp(2N)$ symmetry, the $Sp(2N)/U(N)$ models, flow
to $Sp(2N)_1$.

\end{abstract}

\section{Introduction}

One very interesting property of field theories is that they can have
critical points which are completely unseen in standard weak-coupling
perturbation theory. To prove the existence of these fixed points
requires using alternative perturbative methods (such as the large-$N$
or large-spin expansion), or relying on non-perturbative methods.  One
of the remarkable features of two-dimensional field theories is that
in some cases these non-trivial critical points
can be understood non-perturbatively.

A famous example arose in the study of integer
and half-integer spin chains \cite{Haldane}. The spin-$1/2$ Heisenberg
quantum spin chain is exactly solvable, and from Bethe's exact
solution it is known that the spectrum is gapless.  The obvious guess
for the field theory describing the spin chain in the continuum limit
is the sphere sigma model.  This sigma model is an
$SU(2)$-symmetric field theory where the field takes values on a
two-sphere. The two-sphere can be parametrized by three fields
$(v_1,v_2,v_3)$ obeying the constraint $(v_1)^2 +(v_2)^2 +(v_3)^2=1$.
The Euclidean action is
\begin{equation}
\frac{1}{g} \sum_{i=1}^3 \int dx dy\, (\partial_\mu v_i)^2
\label{o3}
\end{equation}
This field theory, however, is not the correct continuum description
of the spin chain, because it is
also exactly solvable, and its spectrum is gapped \cite{ZZ}. It
was proposed that that when this sigma model is modified by adding
an extra term called
the theta term, the spectrum can become gapless \cite{Haldane}.  The
theta term is inherently non-perturbative: it does not change the beta
function derived near the trivial fixed point at all. 
Nevertheless, the physics can be dramatically different from 
when $\theta=0$ \cite{oblique}.
In two dimensions, this can
result in a non-trivial fixed point. Not only does the spectrum
become gapless at the non-trivial critical point, but in addition,
charge fractionalization occurs: the charges of the quasiparticles of
the theory are fractions of the charge of the fields in the action.

To define the theta term,
field configurations are required to go to a constant at
spatial infinity, so the spatial coordinates ($x,y$) are effectively that
of  a
sphere.  Since the field takes values on a sphere as well, the field is
therefore a map from the sphere to a sphere. An important
characteristic of such maps is that they can have non-trivial
topology: they cannot necessarily be continuously deformed to the
identity map. This is analogous to what happens when a circle is
mapped to a circle (i.e.\ a rubber band wrapped around a pole): you
can do this an integer number of times called the winding number (a
negative winding number corresponds to flipping the rubber band upside
down). It is the same thing for a sphere: a sphere can be wrapped
around a sphere an integer number of times. An example of winding
number 1 is the isomorphism from a point on the spatial sphere to
the same point on the field sphere. The identity map has
winding number 0: it is the map from every point on the spatial
sphere to a single point on the field sphere, 
e.g.\ $(v_1(x,y),v_2(x,y),v_3(x,y))=(1,0,0)$.
Field configurations with non-zero winding number are
usually called instantons. The name comes from viewing one of the
directions as time (in our case, one would think of say $x$ as space
and $y$ as Euclidean time). Since instanton configurations fall off to
a constant at $y=\pm\infty$, the instanton describes a process local
in time and hence ``instant''. Therefore, the field
configurations in the sphere sigma model can be classified by an
integer $n$. This allows a term $$S_\theta = in\theta$$ to be added to
the action, where $\theta$ is an arbitrary parameter. Since $n$ is an
integer, the physics is periodic under shifts of $2\pi$ in $\theta$.
Haldane argued that when $\theta=\pi$, the sphere sigma model flows to
a non-trivial critical point \cite{Haldane}.

Around the same time, a similar proposal arose in
some more general sigma models \cite{Pruisken}.
These models are most easily
formulated by having the field take
values in the coset space $G/H$, where $G$ and $H$ are
Lie groups, with $H$ a subgroup of $G$.  In this language,
the two-sphere is equivalent to $O(3)/O(2)$: while the vector
$(v_1,v_2,v_3)$ can
be rotated by the $O(3)$ symmetry group, it is invariant under the
$O(2)$ subgroup consisting of rotations around its axis.  Thus the
space of distinct three-dimensional fixed-length vectors (the
sphere) is the coset $O(3)/O(2)$.  In order to describe
two-dimensional non-interacting electrons with disorder and a strong
transverse magnetic field, one takes $G=U(2N)$ and $H=U(N)\times U(N)$
\cite{Pruisken}. For $N=1$, one recovers the sphere sigma
model. Pruisken conjectured that in the replica limit $N\to 0$, this
sigma model has a critical point at $\theta=\pi$. This critical point
possibly describes the transitions between integer quantum Hall
plateaus.
Subsequently, a number of conjectures for non-trivial fixed points
in sigma models have been made. For a survey of the applications
of these conjectures to disordered systems, see \cite{nato}.

The purpose of this paper is to generalize this result and prove that
several infinite hierarchies of sigma models have
non-trivial fixed points when $\theta=\pi$. A useful tool is to study
the spectrum and scattering matrix of the particles in an equivalent
$1+1$-dimensional formulation. I will show that when these sigma
models have $\theta=0$, they are gapped, with the spectrum consisting
of massive particles in the symmetric representation of $SU(N)$ and
$O(2P)$ respectively (plus bound states in more general
representations, and kinks with fractional charge in the latter
case)).  When $\theta=\pi$, the spectrum consists of gapless
quasiparticles which are in the fundamental representations (vector,
antisymmetric tensor,\dots) of $SU(N)$ and $O(2P)$ (the latter
including kinks in the spinor representation).

The stable low-energy fixed points for the sphere sigma model and all
the sigma models discussed in this paper are Wess-Zumino-Witten (WZW)
models.  The WZW model for a group $H$ is a sigma model where the
field takes values in $H\times H/H \approx H$.  It also has an extra
term, called the Wess-Zumino term, which has an integer coefficient
$k$. In two dimensions, when $k\ne 0$, the model has a stable
low-energy fixed point \cite{WZW}. The conformal field theory
describing this fixed point is called the $H_k$ WZW model.

One argument for the flow in the sphere sigma model at $\theta=\pi$
goes as follows \cite{Affleck}. First one uses Zamolodchikov's
$c$-theorem, which makes precise the notion that as one follows
renormalization group flows, the number of degrees of freedom goes
down. Zamolodchikov shows that there is a quantity $c$ associated with
any two-dimensional unitary field theory such that $c$ must not
increase along a flow. At a critical point, $c$ is the central charge
of the corresponding conformal field theory \cite{cthm}. At the
trivial fixed point of a sigma model where the manifold is flat, the
central charge is the number of coordinates of the manifold. For the
sphere, this means that $c=2$ at the trivial fixed point.  This if the
sphere sigma model flows to a non-trivial fixed point for
$\theta=\pi$, this fixed point must have $O(3)\approx SU(2)$ symmetry
and must have central charge less than $2$. The only
such unitary conformal
field theories are $SU(2)_k$ for $k<4$. (The central charge
of $SU(2)_k$ is $3k/(k+2)$; in general, the central charge of $H_k$ is
$k\,$dim($H$)$/(k+h)$, where $h$ is the dual Coxeter number of $H$). One
can use the techniques of \cite{KZ} to show that there are relevant
operators at these fixed points, and at $k=2$ or $3$, no symmetry of
the sphere sigma model prevents these relevant operators from being
added to the action \cite{Affleck}. So while it is conceivable that
the sphere sigma model with $\theta=\pi$ could flow near to these
fixed points, these relevant operators would presumably appear in the
action and cause a flow away. However, there is only one relevant
operator (or more precisely, a multiplet corresponding to the WZW
field $w$ itself) for the $SU(2)_1$ theory. The sigma model has a
discrete symmetry $(v_1,v_2,v_3)\to (-v_1,-v_2,-v_3)$
when $\theta=0$ or $\pi$; the winding number $n$
goes to $-n$ under this symmetry, but $\theta=\pi$ and $\theta=-\pi$
are equivalent because of the periodicity in $\theta$. This discrete
symmetry of the sigma model turns into the symmetry $w\to -w$ of the
WZW model. While the operator $\hbox{tr}\, w$ is $SU(2)$ invariant, it
is not invariant under this discrete symmetry. Therefore, this
operator is forbidden from appearing in the effective action. The
operator $(\hbox{tr}\, w)^2$ is irrelevant, so it is consistent for
the sphere sigma model at $\theta=\pi$ to have the $SU(2)_1$ WZW model
as its low-energy fixed point. A variety of arguments involving the
spin chain strongly support this conjecture \cite{Haldane,Affleck}.

An important question is therefore whether the existence of these
non-perturbative fixed points in sigma models at $\theta=\pi$ can be
established definitively.  The fact that the sphere sigma model has a
non-trivial fixed point at $\theta=\pi$ was proven in
\cite{ZZtheta,sausage}. This proof does not involve the spin-$1/2$
chain which motivated the result. Rather, it is a statement about the
non-trivial fixed point in the sphere sigma model at $\theta=\pi$.
This proof
utilizes the integrability of the sphere sigma model at $\theta=0$ and
$\pi$. Integrability means that there are an infinite number of
conserved currents which allow one to find exactly the spectrum of
quasiparticles and their scattering matrix in the corresponding $1+1$
dimensional field theory. The quasiparticles for $\theta=0$ are gapped
and form a triplet under the $SU(2)$ symmetry, while for $\theta=\pi$
they are gapless, and form $SU(2)$ doublets (left- and right-moving)
\cite{ZZ,ZZtheta}.  This is a beautiful example of charge
fractionalization: the fields $(v_1,v_2,v_3)$ form a triplet under the
$SU(2)$ symmetry, but when $\theta=\pi$ the excitations of the system
are doublets. To prove that this is the correct particle spectrum,
first one computes a scattering matrix for these particles which is
consistent with all the symmetries of the theory. From the exact $S$
matrix, the $c$ function can be computed.  It was found that at high
energy $c$ indeed is $2$ as it should be at the trivial fixed point,
while $c=1$ as it should be at the $SU(2)_1$ low-energy fixed point
\cite{ZZtheta}. As an even more detailed check, the free energy at
zero temperature in the presence of a magnetic field was computed for
both $\theta=0$ \cite{Hasen} and $\pi$ \cite{sausage}. The results can
be expanded in a series around the trivial fixed point. One can
identify the ordinary perturbative contributions to this series, and
finds that they are the same for $\theta=0$ and $\pi$, even though the
particles and $S$ matrices are completely different \cite{sausage}.
This is as it
must be: instantons and the $\theta$ term are a boundary effect
and hence cannot be seen in ordinary perturbation theory. One can also
identify the non-perturbative contributions to these series, and see
that they differ. Far away from the trivial fixed point,
non-perturbative contributions can dominate which allow a
non-trivial fixed point to appear when $\theta=\pi$
even though there is none at $\theta=0$.

The purpose of this paper is to
generalize these computations to the 
$SU(N)/SO(N)$ and $O(2P)/O(P)\times O(P)$ sigma models.  
To have any hope of being able to take the replica
limit $N\to 0$, one needs a solution for any $N$.  The $SU(N)/SO(N)$ sigma
model reduces to the sphere sigma model when $N=2$, while $O(2P)/O(P)
\times O(P)$ reduces to two copies of the sphere when $P=2$.  Thus
these sigma models are the generalizations of the sphere sigma model
with $SU(N)$ and $O(2P)$ symmetry respectively.  One difference,
however, is that in general they do not allow a continuous $\theta$
parameter, but instead $\theta$ can only be zero or $\pi$.  I will
show that when $\theta=\pi$, the $SU(N)/SO(N)$ and $O(2P)/O(P)\times
O(P)$ sigma models have stable low-energy fixed points. The
corresponding conformal field theories are the $SU(N)_1$ and $O(2P)_1$
WZW models, respectively.

In section 2 I define the sigma models.
In section 3, I review $S$ matrices in an integrable model with
a global symmetry $G$. In section 4, I discuss the Gross-Neveu
model. This is not a sigma model, but is closely related. These
results are crucial in what follows.  In section 5, I will find the
particles and their scattering matrices for the sigma models. 
In section 6, I will do a substantial check on this
picture by using the exact $S$ matrix to compute the energy in the
presence of a background field. In particular, I check that the
perturbative contributions to this energy are in agreement with
conventional perturbation theory, and that the expansions for
$\theta=0$ and $\pi$ are the same (but the non-perturbative
contributions differ). In section 7, I discuss the low-energy fixed
points at $\theta=\pi$, and flows between them. I also present a
conjecture for the low-energy behavior of the $Sp(2N)/U(N)$ sigma
model. This last section is written to be reasonably self-contained,
so readers interested in the results but not the details
(and willing to trust me) can skip sections 3-6.

\section{The models}

The sigma models discussed in this paper all can be written
conveniently in terms of an matrix field $\Phi$ with
action
\begin{equation}
S= \frac{1}{g}
 \hbox{tr}\, \int d^2x\ \partial^\mu \Phi^\dagger \partial_\mu \Phi
\label{action}
\end{equation}
along with the constraints
\begin{eqnarray}
\Phi^\dagger\Phi &=& I\cr
det(\Phi)&=& \pm 1
\label{restriction}
\end{eqnarray}
where $I$ is the identity matrix. In other words, $\Phi$ is always a
unitary matrix with determinant $\pm 1$, with possibly additional
constraints.
The constraints
(\ref{restriction}) can easily be obtained from theories without
constraints by adding a potentials like
$\lambda\,\hbox{tr}\,(\Phi^\dagger \Phi - I)^2$. When $\lambda$ gets
large, one recovers the restrictions (\ref{restriction}). In theories
with interacting fermions, this often results from introducing a
bosonic field to replace four-fermion interaction terms with Yukawa
terms (interactions between a boson and two fermions). Integrating out
the fermions then results in such potentials and hence the sigma model.

The sigma models discussed in this paper are obtained by
putting additional restrictions of the matrix field $\Phi$.
The two series of models studied both require that
$\Phi$ be symmetric
as well as unitary, so the matrices do not form a group:
multiplying symmetric matrices does not necessarily give
a symmetric matrix. Instead, these spaces is of $G/H$ form.
The $SU(N)/SO(N)$ sigma model is obtained by requiring
that $\Phi$ be a symmetric $N\times N$ unitary matrix with determinant $1$.
The $O(2P)/O(P)\times O(P)$ sigma model is obtained by requiring
that $\Phi$ be a real, symmetric, orthogonal
and traceless $2P\times 2P$ matrix.
In general, in two dimensions a $G/H$ sigma model has
a global symmetry $G$.
The action (\ref{action}) and restriction
(\ref{restriction}) are invariant under the symmetry
\begin{equation}
\Phi \to U \Phi U^T,
\label{symmetry}
\end{equation}
where $U$ is a unitary matrix with determinant one.  This
is the most general symmetry which keeps $\Phi$
symmetric and unitary.
In the $O(2P)/O(P)\times O(P)$ sigma models,
$U$ must be real as well, so $G=O(2P)$.
The field $\Phi$ in this case can
diagonalized with an orthogonal matrix $U$, so
$$\Phi=U\Lambda U^T\qquad\quad \Phi\in O(2P)/O(P)\times O(P),$$
where $U$ is in $O(2P)$, and $\Lambda$ is the matrix with
$P$ values $+1$ and $P$ values $-1$ on the
diagonal.   Different $U$ can result in
the same $\Phi$: the
subgroup leaving $\Phi$ invariant is $H=O(P)\times O(P)$. 
Similarly, field configurations in the
$SU(N)/SO(N)$ sigma model can be written in the form
$$\Phi=U U^T\qquad\quad\Phi\in SU(N)/SO(N)$$
where $U$ is in $SU(N)$. The subgroup $H$ leaving
$\Phi$ invariant is $SO(N)$.
For example, $\Phi=I$ for any real $U$ in $SU(N)$,
i.e.\ if $U$ is in the real subgroup $SO(N)$ of $SU(N)$.  This is
why $H=SO(N)$ here.

These spaces $G/H$ studied here are examples of symmetric spaces.  A
symmetric space $G/H$ has $H$ a maximal subgroup of $G$ (no
subgroup other than $G$ itself contains $H$).  The important property
of a sigma model on a symmetric space is that it contains only one
coupling constant $g$ in the action. In other words, the space $G/H$
preserves its ``shape'' under renormalization, with only the overall
volume changing. The one and two-loop beta functions for all
two-dimensional sigma
models are universal and can be expressed in terms of the curvature of
the field manifold \cite{Friedan}. The effect of renormalization is to
increase the curvature (increase $g$). Even though naively there is no
mass scale in the theory ($g$ is dimensionless), physical
quantities depend on a scale (e.g.
a lattice length or a momentum cutoff)
as a result of short-distance effects. The renormalized coupling depends
on this scale. At short distances, $g$ is small, so the theory is
effectively free: it is a theory of dim$G - $dim$H$ free fields. At
longer distances the theory is interacting: the dim$G - $dim$H$ fields
interact because of the constraint (\ref{restriction}).  In
renormalization-group language, there is an unstable trivial fixed
point at $g=0$, and no non-trivial low-energy fixed point unless one
includes a theta or WZW term.

Both sets of sigma models allow a theta term to be added to the
action.  This is a result long ago proven by mathematicians (for a
discussion accessible to physicists, see \cite{sidney}). In mathematical
language, the question is whether the second homotopy group
$\pi_2(G/H)$ is non-trivial. The second homotopy group is just the
group of winding numbers of maps from the sphere to $G/H$, so for the
sphere it is the integers. The general answer is that $\pi_2(G/H)$ is
the kernel of the embedding of $\pi_1(H)$ into $\pi_1(G)$, where
$\pi_1(H)$ is the group of winding numbers for maps of the circle into
$H$. The rubber band on a pole example means that
$\pi_1(H)$ is the integers when $H$ is
the circle $=U(1)=SO(2)$. The only simple Lie group $H$ for which
$\pi_1$ is nonzero is $SO(N)$, where $\pi_1(SO(N))={\bf Z}_2$ for
$N\ge 3$ and ${\bf Z}$ for $N$=2. Thus there are models with integer
winding number, some with just winding number $0$ or $1$, and some
with no instantons at all. Integer winding number means that $\theta$
is continuous and periodic, while a winding number of $0$ or $1$ means
that $\theta$ is just $0$ or $\pi$ (just think of $\theta$ as being
the Fourier partner of $n$).  The $SU(N)/SO(N)$ and $O(2P)/O(P)\times
O(P)$ sigma models therefore have instantons with ${\bf Z}_2$ winding
number, and so $\theta$ can be zero or $\pi$.

\section{Generalities on exact $S$ matrices}

Even though the sigma models are originally defined in two-dimensional
Euclidean space ($x,y$), it is very convenient to continue to real
time $t=iy$.  The reason is
to treat the field theory as a $1+1$ dimensional
particle theory.  A fundamental
property of many field theories is that all states of the
theory can be written in terms of particles.  In other words, the
space of states is combination of one-particle states, usually called
a Fock space. If the theory has a stable non-trivial fixed point at
low energy, then the particles should be massless: the energy is
linearly related to the momentum: $E=|P|$.  If there is no fixed
point, then the particles are massive: $E=\sqrt{P^2+M^2}$ in a
Lorentz-invariant theory.  In condensed-matter physics, the particles
are often called ``quasiparticles'', to emphasize the fact that the
particles may not be the same as the underlying degrees of freedom:
just because a system is made up of electrons does not mean that the
collective excitations are electrons, or even resemble them. This
phenomenon will occur frequently in this paper.

One finds the scattering matrix of an integrable model by utilizing a
variety of constraints: unitarity, crossing symmetry, global
symmetries, the consistency of bound states, and the factorization
(Yang-Baxter) equations.
It is convenient to write the momentum and energy of a particle in
terms of its rapidity $\theta$, defined by $E=m\cosh\theta$,
$P=m\sinh\theta$. Lorentz invariance requires that the two-particle
$S$ matrix depends only on the rapidity difference $\theta_1-\theta_2$ of
the two particles.

The invariance of the $G/H$ sigma model under the Lie-group symmetry
$G$ requires that the $S$ matrices commute with all group
elements. The $S$ matrix can then be conveniently written in terms of
projection operators.  A projection operator ${\cal P}_k$ maps the
tensor product of two representations onto an irreducible
representation labelled by $k$. By definition, these operators
satisfy ${\cal P}_k{\cal P}_l=\delta_{kl}{\cal P}_k$. Requiring
invariance under $G$ means
that the $S$ matrix for a particle in the representation $a$ with one
in a representation $b$ means that the $S$ matrix is of the form
\begin{equation}
S^{ab}(\theta) = \sum_k f^{ab}_k (\theta) {\cal P}_k
\label{decomp}
\end{equation}
where $\theta\equiv\theta_a-\theta_b$ is the difference of the
rapidities, and the $f^{ab}_k$ are as of yet unknown functions. The
sum on the right-hand side is over all representations $k$ which
appear in the tensor product of $a$ and $b$; of course $\sum_k {\cal
P}_k=1$. In an integrable theory, the functions $f^{ab}_k(\theta)$
are determined up to an overall function by requiring that they
satisfy the Yang-Baxter equation. This stems from the requirement that
the $S$-matrix be factorizable: the multiparticle scattering amplitudes
factorize into a product of two-particle ones. There are two possible
ways of factorizing the three-particle amplitude into two-particle
ones; the requirement that they give the same answer is the
Yang-Baxter equation. There have been hundreds of papers discussing
how to solve this equation, so I will not review this here. For a
detailed discussion relevant to the sigma models here, see e.g.\
\cite{KT,OGW,Mackay}. Solutions arising in the sigma models will
be given below.

To obtain the overall function not given by the Yang-Baxter equation,
one needs to require that the $S$ matrix be unitary, and that it obey
crossing symmetry. With the standard assumption that the amplitude is
real for $\theta$ imaginary, the unitarity relation
$S^\dagger(\theta)S(\theta)=I$ implies $S(\theta)S(-\theta)=I$. The
latter is more useful because it is a functional relation which can be
continued throughout the complex $\theta$ plane.  Crossing symmetry is
familiar from field theory, where rotating Feynman diagrams by $90^o$
relates scattering of particles $a_i$ and $b_j$ to the scattering of
the antiparticle $\bar a_i$ with $b_j$.  In $S$ matrix form, it says
that
\begin{equation}
S^{ab}(i\pi -\theta)= {\bf C}^a S^{\bar{a} b}(\theta) {\bf C}^a,
\label{crossing}
\end{equation}
where ${\bf C}^a$ is the charge-conjugation operator acting on the
states in representation $a$.

Multiplying any $S$ matrix by function $F(\theta)$ which satisfies
$F(\theta)F(-\theta)=1$ and $F(i\pi-\theta)=F(\theta)$ will give an
$S$ matrix still obeying the Yang-Baxter equation, crossing and
unitarity (this is called the CDD ambiguity).  To determine
$F(\theta)$ uniquely, one ultimately needs to verify that the $S$
matrix gives the correct results for the free energy. How to do this
will be described in the following sections. Before doing this calculation,
one must make sure an additional criterion holds:
the poles of the $S$ matrix are
consistent with the bound-state spectrum of the theory. The $S$
matrices of bound states are related to those of the constituents by
the bootstrap relation, which can be formulated as follows
\cite{KT,OGW}. Poles of $S^{ab}$ matrix elements at some value
$\theta=\theta_{ab}$ with $\theta_{ab}$ imaginary and in the
``physical strip''
$0<Im(\theta_{ab})<\pi$ are usually associated with bound
states. Each of the functions $f^{ab}_k(\theta)$ has a residue at this
pole $R_k$. Then there are bound states in representation $k$ if
$R_k\eta_k<0$, where $\eta_k=\pm 1$ is the parity of states in
representation $k$ \cite{Kar}
(poles which do not correspond to bound states give bound states
in the process obtained by crossing). These bound states $(ab)$ have mass
$$(m_{(ab)})^2= (m_a)^2 + (m_b)^2 + 2m_am_b \cosh \theta_{ab}$$
The bound states are not in a
irreducible representation of $G$ if more than one of the residues $R_k$ is
non-zero. 
Then the $S$ matrix for scattering the bound state
$(ab)$ from another particle $c$ is given by
\begin{eqnarray}
S^{(ab)c}=\left(\sum_k \sqrt{|R_k|} {\cal P}_k\right)  S^{ac}\big(
\theta+i\frac{\theta_{ab}}{2}\big)
S^{bc}\big(\theta-i\frac{\theta_{ab}}{2}\big)
\left(\sum_k\frac{1}{\sqrt{|R_k|}}{\cal P}_k\right),
\label{bootstrap}
\end{eqnarray}
where the projection operators ${\cal P}_k$ act on the states in
representations $a,b$. Note that the matrices in (\ref{bootstrap}) are
not all acting on the same spaces, so this relation is to be
understood as multiplying the appropriate elements (not matrix
multiplication).

All of the above considerations apply to massless particles as well,
with a few modifications and generalizations \cite{ZZtheta,FS}.  The
theories being studied are along a flow into their stable low-energy
fixed point. Even though the quasiparticles
are gapless, there is still a mass scale $M$ describing the crossover:
$M\to 0$ is the unstable high-energy fixed point, while $M\to\infty$
is the stable low-energy one.  The rapidity variable for a massless particle
is then defined via $E=me^\theta$, $P=me^{\theta}$ for right movers
and $E=me^{-\theta}$, $P=-me^{-\theta}$ for left movers, where $m$ is not
really the mass of the particle but is proportional to $M$ (different
particles can and do have different values of $m$). The $S$ matrix
still depends only on rapidity differences. Note in particular that
the $S$ matrices for two left movers $S_{LL}$ or two right movers
$S_{RR}$ depend on the ratio $E_1/E_2$ of the two particles, and so do
not depend on the mass scale $M$ at all. These are determined solely
by properties of the low-energy fixed point. On the other hand,
$S_{LR}$ depends on the physics of how one flows into this fixed
point. Massless particles still can have bound states like massive
ones. These bound states show up as poles in $S_{LL}$ and $S_{RR}$,
but it does not seem possible to have poles in $S_{LR}$ in the
physical strip \cite{ZZtheta}.

\section{The Gross-Neveu models}

Before discussing the particle content of the sigma models, it is
useful to first discuss the Gross-Neveu models. These are not sigma
models, but are closely related: they are asymptotically-free field
theories with a global symmetry $G$. For groups $O(N)$ and $U(N)$ they
were originally formulated in terms of fermions with a four-fermion
interaction. For $O(N)$, the action was first written in terms of $N$
Majorana (real) fermions with a four-fermion interaction: \cite{GN}
\begin{equation}
S = \int \sum_{j=1}^{N} \left(\psi_L^j i\partial_L \psi_L^j + \psi_R^j
i\partial_R \psi_R^j\right) + g \left(\sum_{j=1}^N \psi_L^j
\psi_R^j\right)^2.
\label{ongn}
\end{equation}
For $SU(N)$ the action can also be written in terms of fermions (this
is sometimes known as the chiral Gross-Neveu model \cite{natan}), but
for general $G$ the Gross-Neveu models are most easily defined in
terms of the $G$ Wess-Zumino-Witten model at level $1$. The WZW model
has a local symmetry $G_L \times G_R$ generated by chiral currents
$j^a_L$ and $j^a_R$ (the explicit definitions are given in section 7
below). The action of the $G$ Gross-Neveu model is then a perturbation
of the $G_1$ WZW action:
\begin{equation}
S=S(G_1)+g\sum_a \int j_L^a j_R^a
\label{GNaction}
\end{equation}
The coupling $g$ is naively dimensionless, but it has
lowest-order beta function
$$\beta(g) \propto g^2,$$
where the constant of proportionality is negative. This means that for
positive $g$ the coupling is marginally relevant \cite{GN}. In
other words, the WZW fixed point in (\ref{GNaction}) is unstable and
(\ref{ongn}) defines a massive field theory called the Gross-Neveu
model.  For negative $g$, the coupling is marginally irrelevant
and the WZW fixed point is stable, a fact which will become
important for sigma models with $\theta=\pi$.

As one would expect, there are particles in the $O(N)$ Gross-Neveu
model with the same quantum numbers as the fermions in
(\ref{ongn}). These are in the defining ($N$-dimensional vector)
representation of $O(N)$.  For general $G$, this is also true: there
are particles in the defining representation of $G$.  However, there
is more. For $O(N)$, there are kinks as well, as follows from a
semi-classical analysis of the action (\ref{ongn}). These particles
are in the spinor representations of $O(N)$ \cite{witten}. Thus the
$O(N)$ Gross-Neveu model exhibits charge fractionalization.  For even
$N$, there are two spinor representations, of dimension $2^{N/2 -1}$;
for odd $N$ there is one spinor representation, of dimension
$2^{(N-1)/2}$. However, there still more quasiparticles,
which are bound states of the fermions.
As will follow from studying the bound-state
structure of the exact $S$ matrix,
there are particles in all of the
fundamental representations. The fundamental representations for any
Lie group are the representations with
highest weight vectors $\mu^i$ whose inner product with the
simple root vectors $\alpha^j$ obeys $\mu^i \cdot\alpha^j =
\delta^{ij}$ \cite{cahn}. There is one fundamental representation
corresponding to each node on the Dynkin diagram.  The fundamental
representations of $SU(N)$ have Young tableau with only one column:
the defining ($N$-dimensional) representation, the two-index
antisymmetric
tensor, all the way through the $\overline{N}$ representation. Thus for
$G=SU(N)$, there are particles corresponding to any antisymmetric
tensor representation with $j$ indices, $j=1\dots N-1$.
All of these can be considered as bound states
of the $N$-dimensional representation.  For $O(N)$, the fundamental
representations are the antisymmetric tensors and the spinor
representation(s). All of these can be obtained as bound states of the
spinor(s).

The Gross-Neveu models are integrable, so their exact $S$ matrices
can be derived.
For $SU(N)$, the tensor product of two vector representations contains
the symmetric tensor and the antisymmetric tensor. Thus the $S$ matrix
for two particles in the vector representation is of the form
\begin{equation}
S^{VV}= f^{VV}_S{\cal P}_S + f^{VV}_A {\cal P}_A,
\label{vv}
\end{equation}
 where the projection
operators on the two-index symmetric
($S$) and antisymmetric ($A$) representations are
explicitly
\begin{eqnarray}
\nonumber
{\cal P}_S (a_i(\theta_a) b_j (\theta_b)) &=& 
\frac{1}{2} \left(a_i (\theta_a) b_j (\theta_b) +
a_j (\theta_a) b_i (\theta_b)\right)\\
{\cal P}_A (a_i(\theta_a) b_j (\theta_b)) &=& 
\frac{1}{2} \left(a_i (\theta_a) b_j (\theta_b) -
a_j (\theta_a) b_i (\theta_b)\right).
\label{vv2}
\end{eqnarray}
The subscripts in $a_i$ and $b_j$ represent the $i^{\hbox{th}}$ and
$j^{\hbox{th}}$ particles in the vector multiplets, so the $S$ matrix in
(\ref{decomp}) is indeed a matrix.

Requiring that the $S$ matrix obey the Yang-Baxter equation means
that
\begin{equation}
\frac{f^{VV}_A}{f^{VV}_S}=\frac{\theta+2\pi i\Delta}{\theta-2\pi i\Delta}
\label{vv3}
\end{equation}
where $\Delta$ will be determined shortly.
The simplest solution for the overall
function consistent with crossing symmetry and
unitarity for particles in the vector representation of $SU(N)$ is
$f^{VV}_S=S^{VV}_{min}$, where
\begin{equation}
S^{VV}_{min}=
\frac{\Gamma\left(1-\frac{\theta}{2\pi i}\right)
\Gamma\left(\frac{\theta}{2\pi i}+\Delta \right)}
{\Gamma\left(1+\frac{\theta}{2\pi i}\right)
\Gamma\left(-\frac{\theta}{2\pi i}+\Delta\right)}
\label{vv4}
\end{equation}
This is called the minimal solution, because the resulting $S$ matrix
has no poles in the physical strip: a zero in
(\ref{vv4}) cancels the pole in (\ref{vv3}). It is not the unique
solution, because of the CDD ambiguity.

The
Gross-Neveu model has bound states and so the $S$ matrix
must have poles in the physical strip.
This means that $f^{VV}_S= S^{VV}_{min} X(\theta)$, where
$X(\theta)$ contains the poles. If
there are to be bound-state particles with in the antisymmetric
representation of $SU(N)$, but not in the symmetric representation,
then there must be a pole in $f^{VV}_A$ but not in $f^{VV}_S$. The
relation (\ref{vv3}) means the pole must be at $\theta=2\pi i\Delta$ and so
particles in the antisymmetric representation have mass
$m_A/m_V=2\cos\pi\Delta$. To determine $\Delta$, one must ensure that
the bootstrap closes. This means that one does not generate an
infinite number of particles: the poles in all the $S$ matrices
correspond to a finite number of particles. For $SU(N)$, the bootstrap
closes if $\Delta=1/N$. For general groups, the Gross-Neveu bootstrap
closes if $\Delta=1/h$, where $h$ is the dual Coxeter number.  In the
rest of this paper, $\Delta=1/h$, with $h$ appropriate for the case at
hand. Putting (\ref{vv},\ref{vv2},\ref{vv3},\ref{vv4}) and the
bootstrap together means that the $S$ matrix for the scattering of
vector particles in the Gross-Neveu model is \cite{natan}
\begin{equation}
S_{GN}^{VV}(\theta)=X(\theta) S^{VV}_{min}(\theta)
\left({\cal P}_S + \frac{\theta+2i\pi\Delta}{\theta-2i\pi\Delta}
\, {\cal P}_A \right)
\label{vvgn}
\end{equation}
where
\begin{equation}
X(\theta)=\frac{\sinh\big((\theta + 2\pi i\Delta)/2\big)} {\sinh\big(
(\theta -
2\pi i\Delta)/2 \big) },
\label{vvpole}
\end{equation}
The extra factor $X(\theta)$ indeed reinstates the pole at
$\theta=2\pi i\Delta$ canceled by $S_{min}$.
Using this bootstrap to compute the $S$ matrices of
the bound states, one finds for example that the scattering matrix
$S^{AV}$ for a particle in the vector representation with one in the
antisymmetric representation has a pole at $\theta=3\pi i\Delta$, leading
to particles with mass $m_3/m_V
=\sinh(3\Delta\pi)/\sinh(\Delta\pi)$. These particles are in the
representation with highest weight $\mu_3$ (the Young tableau with 3
boxes and one column, or equivalently the three-index antisymmetric
tensor).

One keeps repeating this bootstrap procedure and finds particles in
all the antisymmetric representations, ending up with the particles in
the conjugate representation $\overline{N}$ representation arising
from the bound state of $N-1$ vector particles. This provides a
substantial check on the $S$ matrix, because the crossing relation
(\ref{crossing}) relates $S^{VV}$ to $S^{\overline{V} V}$. One can indeed
check that the $S$ matrices built up from (\ref{vvgn}) satisfy this
relation. For the $SU(N)$ Gross-Neveu model, there is one copy of each
fundamental representation \cite{sungn}.  The dimension of the
antisymmetric tensor with $j$ indices ($j$ boxes in the one-column
Young tableau) is $N!/j!(N-j)!$, so there are $2^N -2$ particles in
all.  Each multiplet of particles has mass $M\sin \pi j/N$, where $M$
is an overall scale related to the coupling constant $g$. The particles
in the $N-j$ representation are the antiparticles of those
in the $j$ representation. One can
check that there are no additional particles by computing the energy
in a background field as done in subsequent sections \cite{Balog}, or
by computing the free energy at non-zero temperature
\cite{Hollo}. Moreover, the $SU(N)$ Gross-Neveu model can be solved
directly using the Bethe ansatz \cite{natan}; the results agree with
those above.

\bigskip\bigskip

The $S$ matrices for the $O(N)$ Gross Neveu model are derived
in the same manner as those with $SU(N)$ symmetry, but there are a
number of additional complications.

The scattering
of particles both in the vector representation of $O(N)$ is of the form
\begin{equation}
{\cal S}^{VV}= {\cal F}^{VV}_S{\cal P}_S + {\cal F}^{VV}_A {\cal P}_A
+{\cal F}^{VV}_0 {\cal P}_0,
\label{ovv}
\end{equation}
 where the projection
operators on the symmetric
($S$), antisymmetric ($A$) and singlet ($0$) representations are
explicitly
\begin{eqnarray}
\nonumber
{\cal P}_S (a_i b_j ) &=& 
\frac{1}{2} \left(a_i  b_j  +
a_j  b_i \right)
-\frac{1}{N} \delta_{ij} \sum_{k=1}^N a_k b_k\\
\nonumber
{\cal P}_A (a_i b_j ) &=& 
\frac{1}{2} \left(a_i  b_j  -
a_j  b_i \right)\\
{\cal P}_0 (a_i b_j )
&=& \frac{1}{N} \delta_{ij} \sum_{k=1}^N a_k b_k
\label{ovv2}
\end{eqnarray}
where I have suppressed the $\theta$ dependence.
For example, the scattering process $a_1b_1\to a_1b_1$ has $S$ matrix
element $((N-2){\cal F}^{VV}_S+N{\cal F}^{VV}_A + 2{\cal
F}^{VV}_0)/2N$, while the process $a_1b_1\to a_2b_2$ has element
$({\cal F}^{VV}_0-{\cal F}^{VV}_S)/N$.  The extra term in (\ref{ovv})
as compared to (\ref{vv}) stems from the fact that the trace
$\delta_{ij}a_i b_j$ is an $SO(N)$ invariant.

Requiring that the $S$ matrix obey the Yang-Baxter equation means
that  \cite{ZZ}
\begin{eqnarray}
\frac{{\cal F}^{VV}_A}{{\cal F}^{VV}_S}&=&
\frac{\theta+2\pi i\Delta}{\theta-2\pi i \Delta}.\\
\frac{{\cal F}^{VV}_0}{{\cal F}^{VV}_S}&=&
\frac{\theta+2\pi i \Delta}{\theta- 2\pi i \Delta}\,
\frac{\theta+i\pi}{\theta-i\pi}.
\label{ovv3}
\end{eqnarray}
The minimal solution (no poles in the physical strip)
for the overall
function for $O(N)$ with particles in the vector is
${\cal F}^{VV}_S={\cal S}^{VV}_{min}$, where
\begin{equation}
{\cal S}^{VV}_{min}(\theta)=
\frac{\Gamma\left(1-\frac{\theta}{2\pi i}\right)
\Gamma\left(\frac{1}{2}+\frac{\theta}{2\pi i}\right)
\Gamma\left(\frac{\theta}{2\pi i}+\Delta\right)
\Gamma\left(\frac{1}{2} -\frac{\theta}{2\pi i}+\Delta\right)
}
{\Gamma\left(1+\frac{\theta}{2\pi i}\right)
\Gamma\left(\frac{1}{2}-\frac{\theta}{2\pi i}\right)
\Gamma\left(-\frac{ \theta}{2\pi i}+\Delta\right)
\Gamma\left(\frac{1}{2} + \frac{\theta}{2\pi i} + \Delta\right)
}
\label{ovv4}
\end{equation}
The $S$ matrix with $f^{VV}_S= {\cal
S}^{VV}_{min}$ defined by (\ref{ovv},\ref{ovv2},\ref{ovv3},\ref{ovv4})
is the $S$ matrix for the $O(N)/O(N-1)$ sigma model \cite{ZZ}.
The $S$ matrix does not contain any poles
because the $O(N)/O(N-1)$ model has no particles in any
representations other than the vector, and hence no bound states. 

The
$O(N)$ Gross-Neveu model has bound states. Just like in $SU(N)$, there are
particles in all the antisymmetric representations, whose
$S$ matrices follow from the bootstrap. These
representations are self-conjugate: there is no $\bar N$
representation. That means that both the bound-state ($s$-channel)
pole at $\theta=2\pi i\Delta$ and the crossed pole ($t$-channel)
channel at 
$\theta=i\pi - 2\pi i\Delta$
appear in ${\cal S}_{GN}^{VV}$, yielding
\begin{equation}
{\cal S}_{GN}^{VV}(\theta)=X(\theta) X(i\pi-\theta)
{\cal S}^{VV}_{min}(\theta)
\left({\cal P}_S + \frac{\theta+2\pi i\Delta}{\theta-2\pi i \Delta}\,
 {\cal P}_A
+\frac{\theta+2\pi i\Delta}{\theta-2\pi i\Delta}\,
\frac{\theta+i\pi}{\theta-i\pi}{\cal P}_0\right)
\label{ovvgn}
\end{equation}
where $X(\theta)$ is as in (\ref{vvpole}) with $\Delta=1/h$, with the
dual Coxeter number here $h=N-2$. This value of $\Delta$ ensures the
bootstrap closes.
The particles obtained by fusing the vector particle $j$ times
therefore have mass $m_{j}/m_{V}=2\sin(j\pi\Delta)/\sin(\pi\Delta)$.
One difference between the $O(N)$ and $SU(N)$ Gross-Neveu models is
apparent in (\ref{ovvgn}): if there is a pole in the antisymmetric
channel at $\theta=2\pi i\Delta$, there is also one in the singlet
channel. This means that there are extra bound states in the $O(N)$
Gross-Neveu model. For example, there are $N(N-1)/2 +1$ particles with
mass $m_2$: $N(N-1)/2$ of them transform in the two-index
antisymmetric representation of $O(N)$, while one is a singlet under
$O(N)$. In general, at mass $m_j$ there are particles in all the
$k$-index completely antisymmetric representations $k=0,2,\dots j$ for
even $j$ and $k=1,3,\dots j$ for odd $j$ \cite{KT}. Despite this
additional complication, one can in principle obtain the $S$ matrices for all
of these states by using fusion. This was done explicitly
for $k=2$ in \cite{Mackay}.

The $S$ matrix for the kinks in the spinor representation in
the $O(N)$ Gross-Neveu model is
given in \cite{KT,OGW,meGN}. It can be written out explicitly, but
since the answer is somewhat complicated I will not give it here. It
is derived by using the fact that kink-kink bound states contain the
particles in the antisymmetric representations as bound states; in
fact is a sort of reverse bootstrap. The mass of the spinor states is
given by $m_{spinor}/m_v=1/\sin(\pi\Delta)$.

In general, the particles of the $G$ Gross-Neveu theory are in
fundamental representations of $G$. Their masses and multiplicities
are known from considering the bound-state properties of the exact $S$
matrix.

\section{The particles and their scattering matrices}

In this section I give the particle content and exact scattering
matrices of the $SU(N)/SO(N)$ and $O(2P)/O(P)\times O(P)$ sigma
models. In the next section, I will use the exact $S$ matrix to
compute the energy in a background field.

The representations of the global symmetries of sigma models are
somewhat more elaborate than that in the Gross-Neveu model.  The
results are different for each model, and also change dramatically if
certain extra terms are added to the action (\ref{action}).

\subsection{$SU(N)/SO(N)$ sigma model}

For $SU(N)/SO(N)$, the action is written in terms of a symmetric
unitary matrix field $\Phi$. Thus as opposed to the Gross-Neveu model,
one expects there to be particles forming the symmetric representation of
$SU(N)$, which has dimension $N(N+1)/2$.
Indeed, it was long ago established that for $N=2$ (the sphere
sigma model), the particles are in the symmetric (triplet)
representation of $SU(2)$ \cite{ZZ}.
For any $N$, there are non-trivial non-local conserved currents in
this sigma model \cite{Abdalla}. These conserved currents are
consistent with particles in the symmetric representation, and also with
bound states in all representations of $SU(N)$ with Young tableau with
two columns and rectangular (i.e.\ the same number of boxes in each of
the columns). In group-theory language, these are representations with
highest weight $2\mu^j$ (the fundamental representations
arising in the Gross-Neveu model have highest weight $\mu^j$).

Building the bound-state $S$ matrices with the bootstrap relation
(\ref{bootstrap}) is one example of a general procedure called fusion
\cite{KRS}.  Fusion exploits the fact that the $S$ matrix can be
written as the sum of projection operators to find new solutions of
the Yang-Baxter equation. It is useful for things other than just the bound
states in a given theory.
The solution of the Yang-Baxter equation $S^{SS}$ for particles are in the
symmetric representation can also be obtained from $S^{VV}$. The solution
is obtained from
(\ref{bootstrap}) like before, using instead the pole at
$\theta=-2\pi i\Delta$. 
The tensor product of two
symmetric representations contains three representations with Young tableaux
$$
\begin{picture}(300,30)(0,10)
\put(30,30){\line(1,0){40}}
\put(30,20){\line(1,0){40}}
\put(30,20){\line(0,1){10}}
\put(40,20){\line(0,1){10}}
\put(50,20){\line(0,1){10}}
\put(60,20){\line(0,1){10}}
\put(70,20){\line(0,1){10}}

%\put(170,10){\line(1,0){10}}
%\put(170,20){\line(1,0){10}}
%\put(170,30){\line(1,0){20}}
%\put(170,40){\line(1,0){20}}
%\put(170,10){\line(0,1){30}}
%\put(180,10){\line(0,1){30}}
%\put(190,30){\line(0,1){10}}

\put(170,15){\line(1,0){10}}
\put(170,25){\line(1,0){30}}
\put(170,35){\line(1,0){30}}
\put(170,15){\line(0,1){20}}
\put(180,15){\line(0,1){20}}
\put(190,25){\line(0,1){10}}
\put(200,25){\line(0,1){10}}

\put(270,15){\line(1,0){20}}
\put(270,25){\line(1,0){20}}
\put(270,35){\line(1,0){20}}
\put(270,15){\line(0,1){20}}
\put(280,15){\line(0,1){20}}
\put(290,15){\line(0,1){20}}

\put(-3,23){$4\mu_1 = $}
\put(110,23){$2\mu_1+\mu_2=$}
\put(230,23){$2\mu_2=$}
\end{picture}
$$
It is convenient to label particles in the symmetric representation
with two indices and the constraint $a_{ij}\equiv a_{ji}$. The projection
operators are then explicitly
\begin{eqnarray}
\nonumber
{\cal P}_{4\mu_1}(a_{ij}b_{kl})&=&\frac{1}{6}
\left(a_{ij}b_{kl}+a_{ik}b_{jl}+a_{il}b_{jk}+a_{kl}b_{ij}
+a_{jk}b_{il}+a_{jl}b_{ik}\right)\\
\nonumber
{\cal P}_{2\mu_1+\mu_2}(a_{ij}b_{kl})&=&\frac{1}{2}
\left(a_{ij}b_{kl}-a_{kl}b_{ij}\right)\\
{\cal P}_{2\mu_2}(a_{ij}b_{kl})&=&\frac{1}{6}
\left(2a_{ij}b_{kl}+2a_{kl}b_{ij}-a_{ik}b_{jl}-a_{il}b_{jk}
-a_{jk}b_{il}-a_{jl}b_{ik}\right)
\end{eqnarray}
This means, for example, that the $S$ matrix element for scattering
an initial state of $a_{12}$ and $b_{13}$ to a final state
of $a_{12}$ and $b_{13}$ is
$(2f^{SS}_{4\mu_1}+3f^{SS}_{2\mu_1+\mu_2}+f^{SS}_{2\mu_2})/6$, while
the element for scattering the same two particles and getting $a_{11}$
and $b_{23}$ in the final state is
$(f^{SS}_{4\mu_1}-f^{SS}_{2\mu_2})/6$.

This enables us to find the $S$ matrix for
the particles in the $SU(N)/SO(N)$ sigma model at $\theta=0$
in the symmetric
representation:
\begin{eqnarray}
S^{SS}= X(\theta) S^{SS}_{min}\,(\theta)\left(
{\cal P}_{4\mu_1} + \frac{\theta+4\pi i\Delta}
{\theta-4\pi i\Delta}\,{\cal P}_{2\mu_1+\mu_2} + \frac{\theta+2\pi i\Delta}
{\theta-2\pi i\Delta}\,\,
\frac{\theta+4\pi i\Delta}{\theta-4\pi i\Delta}\,
{\cal P}_{2\mu_2}\right).
\label{SSsym}
\end{eqnarray}
where $X(\theta)$ is the same as in the Gross-Neveu model.
The overall function $S_{min}^{SS}$ is determined by using unitarity,
crossing and the bootstrap.
The result is that
\begin{equation}
S^{SS}_{min}(\theta)=\frac{\theta-2\pi i\Delta}{\theta+2\pi i\Delta}
\frac{\Gamma\left(1-\frac{\theta}{2\pi i}\right)
\Gamma\left(\frac{\theta}{2\pi i}+2\Delta\right)}
{\Gamma\left(1+\frac{\theta}{2\pi i}\right)
\Gamma\left(-\frac{\theta}{2\pi i}+2\Delta\right)}.
\label{SSmin}
\end{equation}
The symmetric representation has highest weight $2\mu_1$.
The pole in the $S$ matrix at $\theta=2\pi i \Delta$
results in a bound state
transforming in the representation with highest weight $2\mu_2$, with
mass $m_{2\mu_2}/m_S = 2\cos\pi\Delta$.  In fact, the factor $X$
ensures that the spectrum of bound states is just like that in the
$SU(N)$ Gross-Neveu model. Building the $S$ matrices for the
bound states in the $SU(N)/SO(N)$ sigma model proceeds just like the
$SU(N)$ Gross-Neveu model. For example, the prefactor in (\ref{SSsym})
ensures that $S^{S\bar S}$ determined by the bootstrap is the same as
that following from crossing. To close the bootstrap, $\Delta=1/h$
as before, with $h=N$ for $SU(N)$.
There are particles in all representations with highest weight $2\mu_j$.
Each representation appears just once with
mass is $M\sin\pi j/N$.
For $N=2$, this model reduces to the
$O(3)/O(2)$ model (\ref{o3}), and the $S$ matrix (\ref{SSsym})
becomes that of a massive $O(3)$
triplet, familiar from \cite{ZZ}.

\bigskip\bigskip

The $SU(N)/SO(N)$ sigma model with no $\theta$ term is therefore
an integrable generalization of the sphere sigma model to all $SU(N)$.
I will now show that the behavior at $\theta=\pi$ also generalizes.

To find the particles and scattering matrices when $\theta=\pi$, it is
first necessary to understand the non-trivial fixed
point. Generalizing from $N$=2, it is natural to assume that it is
$SU(N)_1$. In fact, the argument from \cite{Affleck,Affleck2} can be
adapted to show that this is completely consistent with the symmetries
of the problem. 
To apply the consistency argument here, it is first useful
to study the symmetries of the $SU(N)$ Gross-Neveu model.
To avoid confusion with the sigma
model, I denote the field in the Gross-Neveu model as
$w$. With the definition (\ref{GNaction}) as a perturbed WZW model,
$w$ must be an $SU(N)$ matrix. Therefore, 
it transforms under the $SU(N)$ symmetry
as $w\to UwU^\dagger$, where $U$ is an element of $SU(N)$.
The currents also transform as $j\to U j U^\dagger$. In
other words, $w$ and $j$ are in the adjoint representation of
$SU(N)$. The action (\ref{GNaction}) is manifestly symmetric under
the symmetry $SU(N)/{\bf Z}_N$. The ${\bf Z}_N$ is the
center of $SU(N)$, consisting of of matrices $\Omega I$,
where $\Omega$ is a $N$th root of unity,
and $I$ is the identity matrix. The matrices $\Omega I$ commute
with all elements of $SU(N)$. The reason this discrete subgroup
is divided out is that if $U=\Omega I$, $w$ and $j$ are left invariant.
Thus the symmetry acting manifestly on the action is $SU(N)/{\bf Z}_N$,
not $SU(N)$. However, the full symmetry of the Gross-Neveu model
is larger. This can be seen in several ways. First of all,
note that there are particles in the model in the vector representation.
The full $SU(N)$ group acts on the vector representation non-trivially.
Indeed, one can assign an extra ${\bf Z}_N$ charge to the particles of
the model, defined so that the particles in the $j$-index
antisymmetric representation have charge $\Omega^j$. Another way
of seeing this extra charge is by examining
terms which do {\it not} appear in the action (\ref{GNaction}).
The $SU(N)/{\bf Z}_N$-symmetric
operators in the $SU(N)_1$ conformal field theory are $\hbox{tr}\,
w^j$, where
$j=1\dots N-1$. Some of these are relevant operators, and if added to
the action would change the physics considerably. However, they are
forbidden from the action if the model is required to be invariant
under the
symmetry $w\to \Omega w$. This discrete symmetry is not
part of the $SU(N)$ acting on the $w$ field,
but rather makes the full symmetry of the model
${\bf Z}_N \times SU(N)/{\bf Z}_N$. This symmetry of the action
and the full $SU(N)$ of the particle description are completely
consistent with each other if the particles are
kinks in the $w$ field. (They are kinks in the $1+1$-dimensional picture,
vortices in the $2+0$-dimensional picture.) The presence of kinks is
easy to see. The field values $w=I$ and $w=\Omega$ are not
related by a continuous symmetry, so field configurations with
$w(x=-\infty,t)=I$ and $w(x=\infty,t)=\Omega I$ are topologically
stable.
These are the kinks. The extra ${\bf Z}_N$ symmetry is then indeed the
discrete kink charge. This sort of symmetry should be familiar
from the sine-Gordon
model, where the soliton charge is not an explicit symmetry of the
action.

The symmetries of the $SU(N)/SO(N)$ sigma model at $\theta=\pi$
are similar to those in the Gross-Neveu model.
The ${\bf Z}_N$ center acts
non-trivially on the $SU(N)/SO(N)$ matrix $\Phi$, as seen in
(\ref{symmetry}). Thus the symmetry of the sigma model action is
the full $SU(N)$ (or to be precise, for even $N$ it is ${\bf Z}_2
\times SU(N)/{\bf Z}_2$).
Therefore it is
consistent for the low-energy fixed point of the $SU(N)/SO(N)$ sigma
model to be $SU(N)_1$. The reason is the same as in the $SU(N)$
Gross-Neveu model: the full $SU(N)$ symmetry of the action
in terms of $\Phi$ implies a
${\bf Z}_N \times SU(N)/{\bf Z}_N$ symmetry
of the low-energy effective action involving $w$. This forbids
the relevant operators from being added to the $SU(N)_1$
action. This argument shows that the
effective action near the low-energy fixed point contains only
$SU(N)$-invariant irrelevant operators. The operator $j_L j_R$ is
$SU(N)$ invariant and of dimension $2$. Thus the effective action here
is like the Gross-Neveu action (\ref{GNaction}), except here the
coupling $g$ is negative so that the perturbing operator is
irrelevant. The coupling $g$ is related to the scale $M$; when
$M\to \infty$, $g\to 0$ from below, and the model reaches the
low-energy fixed point.

The form of the effective action near
the low-energy fixed point gives the $S$ matrices $S_{LL}$
and $S_{RR}$ immediately, because as explained above they are
independent of $M$ and therefore follow solely from $SU(N)_1$. They
must be in fact those of the massless limit of the Gross-Neveu
model. Written as functions of rapidity they are
\begin{equation}
S^{ab}_{LL}(\theta)=S^{ab}_{RR}(\theta)=S^{ab}_{GN}(\theta).
\label{sunLL}
\end{equation}
Thus the spectrum of these gapless particles is the same as that of
the massive particles of the Gross-Neveu model: there are particles in
any antisymmetric representation, a left-moving and right-moving set
for each representation. Even though the particles here are gapless,
the mass ratios of the Gross-Neveu model still appear in the
definition of rapidity. For example, a right mover in the
representation
$\mu^j$ has energy $E=m_j e^\theta$. The charge has fractionalized, as
compared to $\theta=0$.

For $N$=2, the $S$ matrix $S_{LR}(\theta)$ as a function of rapidity
is the same as $S_{LL}(\theta)$ \cite{ZZtheta}. This $S$ matrix
satisfies the Yang-Baxter equation, and has the appropriate $SU(2)$
(not $SU(2)_L\times SU(2)_R$) symmetry. However, the situation is
not quite as simple for general $N$,
because there are poles in $S_{LL}$ and $S_{RR}$
in the physical strip, resulting in the bound states. These
poles are forbidden in
$S_{LR}$. However, it is easy to remove these and still have a sensible
$S$ matrix.  The bound-state $S$ matrices $S_{GN}^{ab}$ are labeled
with $a,b=1\dots N-1$ corresponding to the antisymmetric
representations with $a$ and $b$ indices. For example $S^{11}\equiv
S^{VV}$ and $S^{21}\equiv S^{AV}$. The unwanted
poles in $S^{ab}$
all arise from the factor $X^{11}(\theta)\equiv X(\theta)$
and its fusions. The matrix $S^{ab}_{GN}$ contains the overall
function
\begin{equation}
X^{ab}(\theta)\equiv \prod_{i=1}^{a}\prod_{j=1}^{b}
 X\left(\theta+[2(i+j-1)-a-b]\pi i \Delta\right).
\label{xab}
\end{equation}
For example, $X^{12}= X(\theta+\pi i\Delta) X(\theta-\pi i\Delta)$.
The prefactor $X^{ab}$ contains all the poles in the
physical strip in $S^{ab}_{GN}$. It is easy to check
that $X^{ab}(\theta)X^{ab}(-\theta)=1$ and that
$X^{b\overline{a}}(\theta)=X^{ab}(i\pi-
\theta)$. Therefore
\begin{equation}
S^{ab}_{LR} (\theta) = S^{ab}_{GN}(\theta)\, /\, X^{ab}(\theta)
\label{sunLR}
\end{equation}
satisfies crossing, unitarity, the Yang-Baxter equation, and has no
poles in the physical strip. Thus this is the $S$ matrix for
left-right scattering in the $SU(N)/SO(N)$ sigma model at
$\theta=\pi$. For $N=2$, $X=1$, and the result reduces to
that in \cite{ZZtheta}.

To prove that this picture is correct, in the next
section I will show that these $S$ matrices give a model
which is $SU(N)_1$ in the low-energy limit, but is the $SU(N)/SO(N)$
sigma model at high energy. Moreover, in \cite{meTBA}
the corresponding $c$
function is calculated, and indeed flows from the $c=(N-1)(N+2)/2$
(the value at the trivial high-energy fixed point in $SU(N)/SO(N)$ to
$c=N-1$ (the central charge for $SU(N)_1$).

\subsection{$O(2P)/O(P)\times O(P)$ sigma model}

As with the Gross-Neveu models, the $O(2P)/O(P)\times O(P)$ sigma
model is a slightly more complicated version of its $SU(N)$ analog,
the $SU(N)/SO(N)$ sigma model. The field $\Phi$ is symmetric and
unitary (and also real and traceless), so again one expects particles
in the symmetric representation of $O(2P)$ (which is $P(2P+1)-1$
dimensional), and its various bound states.
This is what happens for $P=2$, where
the model reduces to two copies of the sphere sigma
model: there are six particles: three in each of the symmetric
representations.
Here, for $P>2$, the
non-local conserved currents have not been found, although
some interesting results for the local currents in the classical
model ($g$ small) have been found \cite{evans2}.

The $S$ matrices for the $O(2P)/O(P)\times O(P)$ model can be
constructed by fusing the $O(2P)$ Gross-Neveu results. The $S$ matrix
for particles in the symmetric representation of $O(2P)$ (highest
weight $2\mu_1$) is of the form \cite{Mackay}
\begin{eqnarray}
{\cal S}^{SS}&&=R(\theta)\left({\cal P}_{4\mu_1}+
\frac{\theta+4\pi i\Delta}{\theta-4\pi i\Delta}
\left({\cal P}_{2\mu_1+\mu_2} +
\frac{\theta+i\pi+2\pi i\Delta}{\theta-i\pi-2\pi i
\Delta}\,
{\cal P}_{2\mu_1}\right)\right.\\
\label{oSS}
&&+\frac{\theta+4\pi i\Delta}{\theta-4\pi i\Delta}\,
\frac{\theta+2\pi i\Delta}{\theta-2\pi i\Delta}
\left.\left({\cal P}_{2\mu_2} +
\frac{\theta+i\pi+2\pi i\Delta}{\theta-i\pi-2\pi i\Delta}
\,{\cal P}_{\mu_2} +
\frac{\theta+i\pi+2\pi i\Delta}{\theta-i\pi-2\pi i\Delta}\,
\frac{\theta+i\pi}{\theta-i\pi}\,{\cal P}_0\right)\right)
\nonumber
\end{eqnarray}
where $\Delta=1/h$ as always, with $h=2P-2$ here.
The minimal solution (no poles in the physical strip) for the prefactor
is
\begin{equation}
{\cal S}^{SS}_{min}(\theta)=\frac{\theta-2\pi i\Delta}{\theta+
2\pi i\Delta}\,
\frac{\Gamma\left(1-\frac{\theta}{2\pi i}\right)
\Gamma\left(\frac{1}{2}+\frac{\theta}{2\pi i}\right)
\Gamma\left(\frac{\theta}{2\pi i}+2\Delta\right)
\Gamma\left(\frac{1}{2}-\frac{\theta}{2\pi i}+2\Delta\right)}
{\Gamma\left(1+\frac{\theta}{2\pi i}\right)
\Gamma\left(\frac{1}{2}-\frac{\theta}{2\pi i}\right)
\Gamma\left(-\frac{\theta}{2\pi i}+2\Delta\right)
\Gamma\left(\frac{1}{2}+\frac{\theta}{2\pi i}+2\Delta\right)}.
\label{oSSmin}
\end{equation}
Like the $O(2P)$ Gross-Neveu model, to get bound states
there must be poles at
$\theta=2\pi i \Delta$ and $\theta=\pi i(1-2\Delta)$. This means that
the prefactor of ${\cal S}^{SS}$ is
\begin{equation}
R(\theta) = X(\theta) X(i\pi -\theta) \, {\cal
S}^{SS}_{min}(\theta).
\label{oA}
\end{equation}
Because the factor $X(\theta)X(i\pi-\theta)$ is the same,
the bootstrap for the $O(2P)/O(P)\times O(P)$ model gives bound states
with the same spectrum as the $O(2P)$ Gross-Neveu model.
Also like the $O(2P)$ Gross-Neveu model, there are
particles which do not follow obviously from the action. These
are in ``double-spinor'' representations, formed by the symmetric
product of two spinor representations. In group theory language they
have highest weight $2\mu_s$, where $\mu_s$ is the highest-weight of a
spinor representation; they are $(2P-1)!/P!(P-1)!$ dimensional. These
particles presumably are kinks like in the Gross-Neveu model, but it
is not clear how to extract this information directly from the
action. The $S$ matrix for these kinks is quite complicated, since
many representations appear in the tensor product of two double-spinor
representations. It can presumably be obtained by fusion
of the spinor $S$ matrices, or by the reverse bootstrap like the
Gross-Neveu kinks. The presence of these particles is confirmed by
studying the free energy at non-zero temperature \cite{meTBA}.

Therefore, in both series of sigma models there are particles in all
representations with highest weight $2\mu^i$.  However,
here there are even more
degeneracies and more representations appearing. Since there is a pole
in (\ref{oSS}) at $\theta=2\pi i\Delta$, then there must be bound
states not only in the representation with highest weight $2\mu_2$,
but also in the singlet and in the antisymmetric representation
(highest weight $\mu_2$).  Thus charge is fractionalized even at
$\theta=0$. For example, in $O(8)$, the vector and spinor
representations are $8$-dimensional, the antisymmetric representation
has dimension $28$, the symmetric and double-spinors dimension $35$,
and the representation with highest weight $2\mu_2$ has dimension
$300$. Thus in the $O(8)$ Gross-Neveu model, there are $8$ particles
in the vector (mass $2M\sin\pi/6=M$) $8$ particles in each of the
spinor representations (mass $M$), and $28+1$ particles of mass
$2M\sin\pi/3=\sqrt{3}M$. Thus there are $53$ stable particles in the
$O(8)$ Gross-Neveu model. In the $O(8)/O(4)\times O(4)$ sigma model,
there are $35$ particles in each of the vector and two double-spinor
representations, all of mass $M$. There are $300+28+1$ particles of
mass $\sqrt{3}M$, giving $434$ particles in all.

\bigskip\bigskip

The behavior when $\theta=\pi$ in the $O(2P)/O(P)\times O(P)$ model is
also reminiscent of the $SU(N)/SO(N)$ model at $\theta=\pi$. The
arguments follow in the same fashion.  The model has the $O(2P)_1$ WZW
model as a stable low-energy fixed point. The $O(2P)_1$ WZW model is
equivalent to $2P$ free Majorana fermions, or equivalently $2P$
decoupled Ising models.
The word ``free'' is slightly deceptive,
because just as in a single 2d Ising model, one can study correlators of
the magnetization or ``twist'' operator, which are highly non-trivial.
The consistency argument is simpler here: the
only relevant $O(2P)$ symmetric operators are the fermion mass and the
magnetization operator; neither is invariant under the
symmetry $\Phi\to -\Phi$ of the sigma model. The $c$-function
is computed in \cite{meTBA}, and indeed flows from $c=P^2$ to $c=P$ as
it must.

The $S$ matrix for the quasiparticles follows from the Gross-Neveu
model. The same arguments applied above to the
$SU(N)/SO(N)$ sigma model at $\theta=\pi$ show here that the massless left- and
right-moving particles have the same spectrum as the $O(2P)$
Gross-Neveu model.  Charges fractionalize: the left- and right-moving
particles are in the fundamental representations of $O(2P)$.  The $S$
matrices for the particles in antisymmetric representations are
\begin{eqnarray}
\label{onLL}
&&{\cal S}^{ab}_{LL}(\theta) = {\cal S}^{ab}_{RR}(\theta)
={\cal S}^{ab}_{GN}(\theta)\\
\label{onLR}
&&{\cal S}^{ab}_{LR}(\theta) = {\cal S}^{ab}_{GN}(\theta)
/(X^{ab}(\theta) X^{ab}(i\pi-\theta))
\end{eqnarray}
For particles in the spinor representations, the poles can easily be
removed as well; see \cite{OGW} for the definition of $X^{ab}$ for the
spinor representations.  Even though the field theory at the
low-energy fixed point is free fermions, the $S$ matrix is
factorizable away from the fixed point only in the basis related to
the Gross-Neveu model.

\section{Matching perturbative expansions}

In this section I compute the energy of these sigma models at zero
temperature in a background magnetic field. This is very useful for
several reasons. First of all, it allows a direct comparison of the
$S$ matrix to perturbative results. This in particular ensures that
the $S$ matrices are correct as written in the last section. For
example, it eliminates the possibility of extra CDD factors and/or
extra
bound states. Second,
because the effect of a $\theta$ term is non-perturbative, the
perturbative expansions for the sigma models at $\theta=0$ and $\pi$
must be the same. Thus even though the $S$ matrices for
$\theta=0$ and $\pi$ are very different, the energy must have the same
perturbative expansion. I verify that this is true for the above $S$
matrices.

The abelian subgroup of the group $G$ is $U(1)^r$, where $r$ is the
rank of the group. Thus a model with a global symmetry $G$ has $r$
conserved charges. These charges can be coupled to a background field,
which is constant in spacetime. In the sigma models
$O(2P)/O(P)\times O(P)$ and $SU(N)/O(N)$ where $\Phi$ is a symmetric
matrix, this means that the Euclidean action (\ref{action}) is
modified to
\begin{equation}
S= \frac{1}{g}
 \hbox{tr}\, \int d^2x\
\left(\partial^\mu \Phi^\dagger - \tilde{A}^{T}\Phi^\dagger - \Phi^\dagger
\tilde{A}^{T}\right)
\left(\partial_\mu \Phi + \tilde{A}\Phi + \Phi \tilde{A}\right)
\label{actionmag}
\end{equation}
where $\tilde{A}$ is a matrix in the Cartan subalgebra of $G$ (the
Cartan subalgebra is comprised of the generators of the abelian
subgroup of $G$). For $SU(N)/SO(N)$, $\tilde{A}$ is diagonal with
entries $(A_1,A_2, \dots A_N)$ and the constraint $\sum_i A_i=0$. For
$O(2P)/O(P)\times O(P)$, the matrix can be written in the form
$\tilde{A}_{jk}=\sum_{l=1}^{P} {\cal A}_l
(\delta_{j,2l-1}\delta_{k,2l}- \delta_{j,2l}\delta_{k,2l-1})$.

Because $\tilde{A}$ has dimensions of mass, the energy depends on the
dimensionless parameters $A_i/M$ or ${\cal A}_i/M$. The strength of
the background field controls the position of the theory on its
renormalization group trajectory and, in particular, in the limit of
large field the theory is driven to the ultraviolet fixed point.  To
be near the UV fixed point, all non-zero $A_i/M$ must be large in
magnitude, so let $A$ be one of the non-zero components. The
perturbative expansion around this fixed point therefore is an
expansion for large $A/M$. The computation of this expansion is a
fairly standard exercise in Feynman diagrams; for details closely
related to the cases at hand, see \cite{Balog}. The only effect of the
sigma model interactions resulting from the non-linear constraints
(\ref{restriction}) to the one-loop energy comes through the running
of the coupling constant. Specifically, the two-loop beta function is
of the form
$$\beta(g) = \mu \frac{\partial}{\partial \mu} g(\mu)= -\beta_1 g^2(\mu) -
\beta_2 g^3(\mu)-\dots$$ where $\beta_1$ and $\beta_2$ are
model-dependent, but universal for these sigma models. Solving this
equation means that
\begin{equation}
\frac{1}{g(A)}= \beta_1 \ln A/\Lambda + \frac{\beta_2}{\beta_1}
\ln(\ln A/\Lambda)) +{\cal O}(1/\ln(A))
\label{forg}
\end{equation}
where the scale $\Lambda$ depends on the perturbative scheme used.
The zero-temperature energy for $SU(N)/SO(N)$ through one loop is
$$
E(A)-E(0)= -\frac{4}{g(A)} \sum_j (A_j)^2 - \frac{1}{2\pi} \sum_{i<j}
(A_i-A_j)^2 \left(\ln|A_i-A_j| - \frac{1}{2}\right) + {\cal O}(g),
$$
with a similar formula for $O(2P)/O(P)\times O(P)$.
For both cases,  this means that at large $A/M$, the energy is of the form
\begin{equation}
E(A)-E(0) \propto A^2(\ln (A/M) +
\frac{\beta_2}{(\beta_1)^2} \ln(\ln(A/M))+\dots)
\label{gse}
\end{equation}
In particular, note that the ratio of the first two terms is universal.
These logarithms are characteristic of an asymptotically free theory,
where the perturbation of the UV fixed point is marginally relevant.

I now explain how to calculate the ground-state energy $E(A)-E(0)$
directly from the $S$-matrix, giving the promised check. In this
picture, the model is treated as a $1+1$ dimensional particle theory
at zero temperature. Turning on the background field has the effect of
changing the one-dimensional quantum ground state. If one were working
in field theory or in a lattice model, this would change the Dirac or
Fermi sea. In the exact $S$-matrix description, a similar thing
happens: the ground state no longer is the empty state --- it has a
sea of real particles. For example, a particle $a_i$ in the vector
representation of $SU(N)$ has its energy shifted by $A_i$ in a
magnetic field. If the total energy is negative, then it is possible
for such a particle to appear in the ground state.  Because the
technicalities are slightly different, I treat the cases of massive
and massless particles separately.

\subsection{Massive particles}

For simplicity, I first treat only the case where
the magnetic field is chosen so that only
one kind of massive particle of mass $m$ and charge $q$ appears in the
ground state. In integrable
models obeying the Yang-Baxter equation, the particles fill levels
like fermions: only one can occupy a given level.  Then the ground
state is made up of particles with rapidities
$$-B < \theta_1 < \theta_2 <\dots \theta_{\cal N} <B$$
for some maximum rapidity $B$.
When the length $L$ of the system is large, these levels are very close
together, so the density of particles $\rho(\theta)$ is defined
so that $\rho(\theta)d\theta$ is
the number of particles with rapidities between $\theta$ and 
$\theta+d\theta$.
The energy density of the ground state is then
\begin{equation}
E(A)-E(0)= {1\over L}\sum_\alpha
 \int_{-B}^{B} d\theta\ \rho(\theta)(-A+m \cosh\theta).
\label{energy}
\end{equation}
I have normalized $A$ so that the charge $q=1$.
The exact $S$-matrix allows us to derive equations for
the $\rho(\theta)$
and $B$. Imposing periodic boundary
conditions on the box of length $L$ requires that the rapidities
$\theta_i$ of the particles in the ground state
all satisfy the quantization conditions
\begin{equation}
e^{im\sinh \theta _i L}\prod_{j\ne i}
S(\theta_i-\theta_j) = 1
\label{quant}
\end{equation}
for all $i$. This is an interacting-model generalization of the
one-particle relation $m\sinh\theta_k=p_k=2\pi n_k/L$ to the case
where the particles in the ground state scatter elastically from each
other with $S$ matrix element $S(\theta)$.  In the large $L$ limit,
taking the derivative of the log of (\ref{quant}) gives an integral
equation for the density $\rho(\theta)$:
\begin{equation}
\rho(\theta)= {mL\over 2\pi} \cosh\theta+
\int_{-B}^{B} d\theta'\rho(\theta')\phi(\theta-\theta'),
\label{forrho}
\end{equation}
where $$\phi(\theta)\equiv -{i\over 2\pi}
{\partial\ln{S(\theta)}\over\partial\theta}.$$
This equation is valid for $|\theta|<B$; for $|\theta|>B$, $\rho(\theta)=0$.
The maximum rapidity $B$ is determined by minimizing the energy equation
(\ref{energy}) with respect to $B$ subject to the constraint (\ref{forrho}).
If the particles were non-interacting, the density would be $mL\cosh\theta$
and
$m\cosh B=A$, but the effect of the interactions is quite substantial.

These equations can be put in a convenient form
by defining the ``dressed'' particle energies $\epsilon(\theta)$ as
\begin{equation}
\epsilon(\theta)=A - m\cosh \theta  +
\int_{-B}^{B} d\theta' \phi(\theta -\theta')\epsilon(\theta').
\label{forep}
\end{equation}
Substituting this into (\ref{energy}) and using (\ref{forrho} yields
\begin{equation}
E(A)-E(0)= -{m\over 2\pi}\int_{-B}^{B}d\theta\ \cosh\theta
\ \epsilon(\theta).
\label{forE}
\end{equation}
In this formulation,  $B$ is a function of $A/m$
determined by the boundary condition
$\epsilon(\pm B)=0$.

Consider the $O(2P)/O(P)\times O(P)$ sigma model, and choose the
magnetic field to be ${\cal A}_1=A$, ${\cal A}_i=0$ for $i\ne 1$.  The
massive particles used above ($a_{ij}=a_{ji}$) are not eigenstates of
the matrix magnetic field $\widetilde{A}$, but one can easily change basis
to
particles which are. This merely requires taking linear combinations
of particles of the same mass. The state which has the largest
eigenvalue of this magnetic field is $d\equiv
a_{11}-a_{22}+2ia_{12}$.
This basis also has the advantage that scattering of $d$
particles amongst themselves is diagonal (as opposed to
the scattering of $a_{11}$ with $b_{11}$, which does not necessarily yield
$a_{11}$ and $b_{11}$ in the final state). The $S$ matrix element for
scattering $d(\theta_1)d(\theta_2)$ to $d(\theta_1) d(\theta_2)$ is
$R(\theta_1-\theta_2)$ as given in (\ref{oA}).  The state $d$ has the
largest eigenvalue of the magnetic field, and is the only state with
this eigenvalue. I therefore make the assumption that it is the only
kind of particle which appears in the ground state. This assumption is
standard in these computations; it can be justified by a careful
treatment of the zero-temperature limit of the finite-temperature
density equations discussed in \cite{meTBA}.  The kernel $\phi$ of the
integral equation (\ref{forep}) is then proportional to the derivative
of $\log R(\theta)$. It is convenient to write this in Fourier space
as
\begin{eqnarray}
\label{phiK}
&&\phi(\theta)= \int_{-\infty}^\infty d\omega\
e^{i\omega\theta} (1-{\cal K}(\omega))\\
&&{\cal K}(\omega)=e^{-\pi \Delta|\omega|} \frac{2\cosh((1-2\Delta)\pi
\omega/2)
\sinh(2\pi\Delta|\omega|)}{\cosh(\pi\omega/2)}
\label{forcalK}
\end{eqnarray}
where $\Delta$ is as always $1/h$, with the dual Coxeter number
$h=2P-2$ for $O(2P)$.

For the $SU(N)/SO(N)$ sigma model, the states $a_{ij}$ are eigenstates
of the magnetic field operator, with eigenvalue $A_{i}+A_{j}$. If we
chose the magnetic field to be $A_1=A$, $A_j=-A/(N-1)$ for $j>1$,
then the
particle $a_{11}$ has maximum charge. Again making the assumption that
this is the only particle in the ground state, the Fourier transform
of the resulting kernel is
$$2 e^{-\pi \Delta|\omega|}\frac{\sinh((1-\Delta)\pi|\omega|)
\sinh(2\pi\Delta\omega)}{\sinh(\pi\omega)}$$
However, it is convenient here to instead choose the fields to be
$A_1=-A_2=A$, $A_j=0$ for $j>2$ (this choice of field
was useful also in the
supersymmetric $CP^n$ model \cite{Evans}) . Then there are two
particles with largest eigenvalue: $a_{11}$ and the antiparticle (in
the $\bar N$ representation) $\bar{a}_{22}$. Thus they both appear in
the ground state. These two particles scatter diagonally among
themselves and each other. The $S$ matrix element $S_1$ for
$a_{11}a_{11}\to a_{11}a_{11}$ and ${\bar a}_{22}{\bar a}_{22} \to
{\bar a}_{22} {\bar a}_{22}$ is
$$S_1 (\theta)= S^{SS}_{min}(\theta)X(\theta).$$
The element $S_2$ for  $a_{11}\bar{a}_{22}\to a_{11}\bar{a}_{22}$
is found easily from (\ref{SSsym}) by using crossing. It is
$$S_2(\theta) = X(i\pi-\theta) S^{SS}_{min}(i\pi-\theta)
\frac{(i\pi-\theta)(i\pi-\theta+\mu)}{(i\pi-\theta-2\mu)(i\pi-\theta-\mu)}$$
Because of the symmetry between the two kinds of particles, their
ground-state densities must be the same. The ground-state energy then
follows from a simple generalization of the above analysis: the
equations (\ref{forep}) and (\ref{forE}) still apply, with
$$\phi(\theta)\equiv -{i\over 2\pi} \frac{\partial}{\partial\theta}
\left(\ln{S_1(\theta)} + \ln{S_2(\theta)}\right)$$
here.  Plugging in
the explicit expressions of the functions yields the useful fact that
$S_1(\theta)S_2(\theta)=R(\theta)$, where $R(\theta)$ is the function
appearing in (\ref{oA}). Thus with these choices of magnetic field,
the $O(2P)/O(P)\times O(P)$ and the $SU(N)/SO(N)$ sigma models can be
treated by using the kernel (\ref{forcalK}): the only difference is the
$\Delta=1/(2P-2)$ in the former and $\Delta=1/N$ in the latter.

The linear integral equation (\ref{forep}) cannot be solved in closed
form. However, there is a generalized Weiner-Hopf technique which
allows the perturbative (and non-perturbative) expansion for large
$A/M$ to be obtained \cite{JNW}. I discuss this technique in the next
subsection. In order to compare the energy in the $\theta=0$ sigma
model with the perturbative computation (\ref{gse}), I can rely on the
results of \cite{Balog}. There the first few terms in the large $A$
expansion for models with kernels like (\ref{forcalK}) are given.  The
technique requires that the kernel ${\cal K}(\omega)$ be factorized
into
\begin{equation}
{\cal K}(\omega)=\frac{1}{{\cal K}_+(\omega){\cal K}_-(\omega)}
\label{factorK}
\end{equation}
where ${\cal K}_+(\omega)$ has no poles or zeroes (and is bounded) in the
upper half plane $Im(\omega)>0$,
while ${\cal K}_-(\omega)={\cal K}_+(-\omega)$ has
no poles or zeroes and is bounded in the upper half plane. Then
\begin{equation}
{\cal K}_+(\omega)=\sqrt{\frac{2i\Delta\omega}{\pi}}
e^{i\omega\Delta\ln(i\omega) +i\mu\omega}
\frac{\Gamma(-2i\Delta\omega)
\Gamma\left(\frac{1}{2}-i(\frac{1}{2}-\Delta)\omega \right)}
{
\Gamma\left(\frac{1}{2} - i\frac{\omega}{2}\right)}.
\end{equation}
where
$$\mu= 2\Delta\ln(2\Delta) +
(\frac{1}{2}-\Delta)\ln(\frac{1}{2}-\Delta) +
\frac{1}{2}\ln 2 - \Delta.$$
The factor $e^{i\omega\mu}$ ensures that ${\cal K}_+$ is bounded
appropriately as $|\omega|\to\infty$ in the upper half plane.
When for small $\xi$, $K_+(i\xi)$ goes
as
$$K_+(i\xi)= \frac{k}{\sqrt{\xi}} e^{s\xi\ln\xi}(1+\dots),$$
the energy at large $A$ is \cite{Balog}
\begin{equation}
E(H)-E(0)= -\frac{k^2}{4}A^2\left[\ln\left(\frac{A}{m}\right) +
(s+\frac{1}{2}) \ln\left(\ln\left(\frac{A}{m}\right)\right) + \dots
\right]
\label{Eexpand}
\end{equation}
This expansion is of the same form as (\ref{gse}). The two are equal
if
\begin{equation}
\frac{\beta_2}{(\beta_1)^2}=s+\frac{1}{2}.
\label{condition}
\end{equation}
The explicit kernel (\ref{forcalK}) yields
$$s=\Delta,$$
while perturbative computations (see e.g. \cite{hikami} and
references within) give for $SU(N)/SO(N)$
$$
\frac{\beta_2}{(\beta_1)^2}=\frac{N+N^2/2}{N^2}
$$
while for $O(2P)/O(P)\times O(P)$ the ratio is
$$
\frac{\beta_2}{(\beta_1)^2}=\frac{2P^2-2P}{(2P-2)^2}
$$
Using the relations $\Delta=1/N$ and $\Delta=1/(2P-2)$ respectively, one
 indeed sees
that that the condition (\ref{condition}) holds. This is a substantial
check, and coupled with the fact that $S$ matrix also gives the
correct central charge, leads me to be completely convinced that the
$S$ matrices discussed above are indeed the correct sigma model $S$
matrices.

\subsection{Comparing the perturbative expansions at $\theta=0$ and $\pi$}

In this subsection I analyze the equations for the ground-state energy
in a magnetic field in more detail. The final result will be that the
{\it entire} large-$A$ perturbative expansions for $\theta=0$ and for
$\theta=\pi$ are identical, but that non-perturbative contributions
differ. This effectively confirms the $S$ matrices and spectrum above,
and the identity of the low-energy fixed points.

The generalized Weiner-Hopf technique is discussed in detail in the
appendix of \cite{JNW}. The equation to be solved is of the form
$$\epsilon(\theta)-\int_{-B}^{B}
\phi(\theta-\theta')\epsilon(\theta')=g(\theta).$$
where $\epsilon(\theta)$ and $g(\theta)$ are both vanishing for
$|\theta|\ge B$, and $\phi(\theta)=\phi(-\theta)$.
Defining the
Fourier transforms $\widetilde{\epsilon}(\omega)$,
$\widetilde{g}(\omega)$ and ${\cal K}(\omega)$ (the latter related to $\phi$
as in (\ref{phiK})) gives
$$\int_{-\infty}^\infty e^{-i\omega\theta}
\left\{\widetilde{\epsilon}(\omega)
{\cal K}(\omega)- \widetilde{g}(\omega)\right\}=0.$$
This equation is valid for $|\theta|<B$. Fourier-transforming this gives
\begin{equation}
\widetilde{\epsilon}(\omega){\cal K}(\omega)-\widetilde{g}(\omega) =
X_+(\omega) e^{i\omega B} + X_-e^{-i\omega B}.
\label{wh1}
\end{equation}
The functions $X_\pm$ arise, roughly speaking, from the analytic
continuation of $\epsilon(\theta)$ to $|\theta|>B$. The extra factors
$e^{i\omega B}$ ensure that $X_+(\omega)$ is analytic in the upper
half plane, while $X_-(\omega)=X_+(-\omega)$ is analytic in the lower
half plane.
%Using the factorization (\ref{factorK}) gives
%\begin{equation}
%\frac{\epsilon_\pm(\omega)}{{\cal K}_\pm(\omega)}
%-g_\pm(\omega) {\cal K}_\mp(\omega)= X_\mp(\omega) G_\mp(\omega)
%+ X_\pm(\omega) {\cal K}_\mp(\omega) e^{\pm 2i\omega B} ,
%\label{crucial}
%\end{equation}
The relation (\ref{wh1})
can be split into two equations, one involving poles
in the upper half plane and the other the lower. To split a given function,
one uses
$$f(\omega)=[f(\omega)]_+ + [f(\omega)]_-$$
where
$$[f(\omega)]_\pm = \pm \frac{1}{2\pi i} \int_{-\infty}^\infty
\frac{f(\omega)}{\omega'-\omega\mp i\delta} d\omega'$$
where $\delta$ is a positive real number tending to zero. The functions
$[f]_\pm$
are analytic in the upper ($+$) and lower ($-$) half planes.
Because $\epsilon(\theta)$ and $g(\theta)$ are zero for
$|\theta|\ge B$, the functions
$$ \epsilon_\pm (\omega)= \widetilde{\epsilon}(\omega) e^{i\omega B}
\qquad\qquad
g_\pm (\omega)= \widetilde{g}(\omega) e^{i\omega B}$$
are similarly analytic in the upper half and lower half planes.
Equations for the functions $X_\pm$ can be derived by
exploiting these analyticity properties and the factorization
(\ref{factorK}). Namely, (\ref{wh1}) implies the equations
\begin{equation}
\frac{\epsilon_\pm(\omega)}{{\cal K}_\pm(\omega)}
=[g_\pm(\omega) {\cal K}_\mp(\omega)]_\pm
+ [X_\pm(\omega) {\cal K}_\mp(\omega) e^{\pm2i\omega B}]_\pm ,
\end{equation}
while the $X_\pm$ are given by
\begin{equation}
X_\pm(\omega){\cal K}_{\pm}(\omega)
+ [X_\mp(\omega) {\cal K}_\pm(\omega) e^{\mp 2i\omega B}]_\pm
=-[g_\mp(\omega) {\cal K}_\pm(\omega)]_\pm
\label{forX}
\end{equation}
Unfortunately, it is not possible to solve the equations for $X_\pm$
explicitly. However, this form does allow the large-$A$ expansion to
be systematically developed.

To study the large $A$ expansion, 
it turns out to be much
easier to consider the more general kernel
\begin{equation}
K(\omega)= \frac{\cosh(\gamma\pi\Delta\omega)}
{\sinh((\gamma+1)\pi\Delta\omega)}
\frac{2\cosh((1-2\Delta)\pi\omega/2)\sinh(2\Delta\pi\omega)}
{\cosh(\pi\omega/2)}.
\label{forK}
\end{equation}
In the limit $\gamma\to\infty$, $K$ goes back to the sigma model
kernel ${\cal K}$. In the $N$=2 case, this deformation corresponds
to deforming the sphere sigma model into the sausage sigma model
\cite{sausage}.
This kernel factorizes as $K=1/(K_+K_-)$, giving
\begin{equation}
{K}_+(\omega)=\sqrt{\frac{2}{\gamma+1}}
e^{i\omega\nu}
\frac{\Gamma\left(\frac{1}{2} - i\Delta\gamma\omega\right) 
\Gamma\left( \frac{1}{2} - i(\frac{1}{2}+\Delta)\omega \right)
\Gamma\left(-2i\Delta\omega\right)}
{\Gamma\left(-i(\gamma+1)\Delta\omega\right) 
\Gamma\left(\frac{1}{2} - i\frac{\omega}{2}\right)}.
\end{equation}
where 
$$\nu=\mu+\gamma\Delta\ln(\gamma/(\gamma+1)) - \Delta\ln\Delta +\Delta.$$
The equations (\ref{forX}) can be written as a single one by exploiting
the relation $X_+(\omega)=X_-(-\omega)$.
Defining 
$$v(\omega)\equiv \frac{X_+(\omega)}{K_+(\omega)} -
i\frac{qK_+(0)}{\omega+i\delta} A -
i\frac{me^B}{2}\,\frac{K_+(i)}{\omega-i},$$ they become
\begin{equation}
v(\omega)=-\frac{iAK_+(0)}{\omega+i\delta} + \frac{ime^{B}}{2}
\frac{K_+(i)}{\omega-i}+\int_{{\cal C}_+} \,
\frac{e^{2\omega i B}}{\omega+\omega'+i\delta}\, \alpha(\omega') v(\omega') 
\frac{d\omega'}{2\pi i}
\label{forv}
\end{equation}
where $\alpha(\omega)$ is defined as
$$\alpha(\omega)\equiv \frac{K_-(\omega)}{K_+(\omega)}$$
and the integration contour circles all singularities on the positive
imaginary axis.
The boundary condition $\epsilon(\pm B)=0$ becomes
\begin{equation}
iAK_-(0)-\frac{ime^{B}}{2}K_-(-i)=
\int_{{\cal C}_+} \,
e^{2\omega i B} \alpha(\omega) v(\omega) 
\frac{d\omega}{2\pi i}
\label{bc}
\end{equation}
while the energy (\ref{forE}) becomes
\begin{equation}
E(A)-E(0)=\frac{me^{B}}{2}K_-(-i)\left[AK_+(0)-\frac{me^{B}}{4}K_+(i)
\int_{{\cal C}_+} \,
\frac{e^{2\omega i B}}{\omega-i}\, \alpha(\omega) v(\omega) 
\frac{d\omega}{2\pi i}\right].
\label{forEii}
\end{equation}
The contour here includes the double pole at $\omega=i$, and the
poles in $\alpha(\omega)$. The function $v(\omega)$ has no poles in
the upper half plane except the explicit one at $\omega =i$.

These equations are convenient for deriving the expansion for large
$A$. The reason is that when $A$ is large, the range of rapidities
allowed in the ground state is large, so $B$ is large (recall that if
the particles were free, $A=m\cosh B$). In this limit the integrals in
(\ref{forv},\ref{bc},\ref{forEii}) are small corrections, and their
effect can be treated iteratively. This iterative expansion is worked
out in detail in \cite{Zsine}.  The first correction to $v(\omega)$
comes from approximating the integral using the leading pieces of
$v(\omega)$ in the integrand. The boundary condition (\ref{bc})
relates $B$ to $A$. Thus the contributions to this integral
come from the pole in $v(\omega)$ at $\omega=i$ and the poles in
$\alpha(\omega)$.  The poles of $\alpha$ are at
$\omega=i(2n-1)/(2\gamma\Delta)$, $\omega=(2n-1)h/(2\Delta +1)$ and
$\omega=in/(2\Delta)$, for $n$ a positive integer. Using the iterative
procedure it is straightforward to see the form of the expansion for
large $A$: it is a power series of the form
\begin{equation}
E(A)= A^2\left(\sum_{j=0}^\infty \alpha_j 
\left(\frac{A}{m}\right)^{j/(\gamma\Delta)}\right)
\left(1+{\cal O}\left(\frac{A}{m}\right)^2\right).
\label{power}
\end{equation}
The first series is a result of the first series of poles in $\alpha$;
the coefficients $\alpha_j$ depend on the residues at
these poles. The order $A^2$ corrections are a result of the pole at
$\omega=i$ and the other poles in $\alpha$. Recall that $\gamma$ is large,
so these contributions are much smaller. The double pole at
$\omega=i$ results in an $A$-independent piece, which is identified as
$E(0)$ \cite{sausage}.

The sigma model result of interest is recovered in the limit
$\gamma\to\infty$. This complicates matters considerably for the power
series in (\ref{power}), since the exponent is going to zero.  This is
precisely what happens for example in the anisotropic Kondo problem as
the anisotropy is tuned away \cite{TW}.  What happens in this limit is
that the power series turns into an expansion in
$\ln(A),\ln(\ln(A)),\dots$. Note from (\ref{forg}) that $1/g \propto
\ln(A)+ \dots$, so the series depends on powers of $g$, while the
order $A^2$ corrections depend on $e^{-const/g}$.  Thus the first
series consists of the perturbative contributions to the free energy,
while the second part is non-perturbative: the latter will never be
seen in standard sigma model perturbation theory in $g$.  The exact
form of the perturbative expansion cannot be displayed in closed form,
but one can build it piece by piece; the first pieces are displayed in
(\ref{Eexpand}).

\bigskip\bigskip The analogous equations for $E(A)$ for the sigma
models with $\theta=\pi$ follow from their $S$ matrices.
I will show that the large $A/M$ expansion
is of the form (\ref{power}), with the {\it identical}
$\alpha_n$. This effectively proves that these $S$ matrices are
those of the sigma model at $\theta=\pi$. The flow to the WZW
model then follows immediately, because the $S$ matrix manifestly
becomes that of the
appropriate WZW model in the large-$M$ limit.

The integral equations for massless particles in a background field
are similar to the ones for massive particles. The massless particles
are left and right moving, and because they are gapless, they begin
filling the sea for arbitrarily field $A$. Therefore, right moving
particles with rapidities $-\infty<\theta<B$ and left moving particles
with rapidities $-B<\theta<\infty$ fill the sea. For one species of
right mover, and one species of left mover, the analysis at the
beginning of this section can be repeated to derive the equations
\begin{eqnarray}
\label{epR}
\epsilon_R(\theta)&= &A - \frac{M}{2} e^\theta + 
\int_{-\infty}^B \phi_1(\theta-\theta')\epsilon_R(\theta')d\theta'
+\int_{-B}^{\infty} \phi_2(\theta-\theta')\epsilon_L(\theta')d\theta'\\
\label{epL}
\epsilon_L(\theta)&= &A - \frac{M}{2} e^{-\theta} + 
\int_{-\infty}^B \phi_2(\theta-\theta')\epsilon_R(\theta')d\theta'
+\int_{-B}^{\infty} \phi_1(\theta-\theta')\epsilon_L(\theta')d\theta'
\end{eqnarray}
valid for $-\infty<\theta<B$ and $-B<\theta<\infty$ respectively,
with the boundary conditions
$$\epsilon_R(B)=\epsilon_L(-B)=0.$$
The kernels are defined as
$$\phi_1(\theta)\equiv -{i\over 2\pi} \frac{\partial}{\partial\theta}
\ln{S_{LL}(\theta)}\qquad  
\phi_2(\theta)\equiv -{i\over 2\pi} \frac{\partial}{\partial\theta}
\ln{S_{LR}(\theta)}
$$
The energy is 
\begin{equation}
E^{(\pi)}(A)-E^{(\pi)}(0) = -\frac{M}{2\pi} \int_{-\infty}^B
e^\theta \epsilon_R(\theta)d\theta'
\end{equation}
where I have used the symmetry
$\epsilon_L(\theta)=\epsilon_R(-\theta)$.

The two equations (\ref{epR},\ref{epL}) can be made into one
by exploiting the
left-right symmetry. In terms of Fourier transforms,
$\widetilde\epsilon_R(\omega)=\widetilde\epsilon_L(-\omega)$, giving
$$
\widetilde{\epsilon}_R(\omega)k_1(\omega)
-\widetilde{\epsilon}_R(-\omega)k_2(\omega)
-\widetilde{g}_R(\omega) =
Y_+(\omega) e^{i\omega B}
$$
where the symmetry $k_2(-\omega)=k_2(\omega)$ is used,
and $g_R(\theta)=A-me^{\theta}/2$.  Note only one
term is need on the right-hand side, because the integral in $\theta$
space runs all the way to $-\infty$. Getting rid of the
$\epsilon_R(-\omega)$ gives
\begin{equation}
\widetilde{\epsilon}_R(\omega)\left(k_1(\omega)-
\frac{(k_2(\omega))^2}{k_1(\omega)}\right)
-\widetilde{g}_R(\omega) -\widetilde{g}_R(-\omega)
\frac{k_2(\omega)}{k_1(\omega)}
=Y_+(\omega) e^{i\omega B} +Y_-(\omega) e^{-i\omega B}
\frac{k_2(\omega)}{k_1(\omega)}
\end{equation}
where as always $Y_-(\omega)=Y_+(-\omega)$.
This is now an equation of the form (\ref{wh1}),
and the same generalized Weiner-Hopf analysis can be applied.
Defining
$$k(\omega)\equiv k_1(\omega) - \frac{(k_2(\omega))^2}{k_1(\omega)}$$
and factorizing it as $k(\omega)=1/(k_+(\omega)k_-(\omega))$ in the usual
way, one finds an equation just like (\ref{forv}), namely
\begin{equation}
v(\omega)=-\frac{iAk_+(0)}{\omega+i\delta}\frac{k_1(0)}{k_2(0)}
 + \frac{ime^{B}}{2} 
\frac{K_+(i)}{\omega-i} \frac{k_1(i)}{k_2(i)}+\int_{{\cal C}_+} \,
\frac{e^{2\omega i B}}{\omega+\omega'+i\delta}\, \beta(\omega') 
v(\omega') 
\frac{d\omega'}{2\pi i}
\label{forvpi}
\end{equation}
where
$$\beta(\omega)=\frac{k_-(\omega)}{k_+(\omega)} 
\frac{k_1(\omega)}{k_2(\omega)}.$$
The boundary conditions and energy follows with the same substitutions.

Now I can show that the sigma models have the same perturbative
expansions at $\theta=0$ and $\pi$.  Using the magnetic field
described in the last subsection, the kernels $\phi_1$ and $\phi_2$
follow simply from the Gross-Neveu $S$ matrix and the relations
(\ref{sunLL},\ref{sunLR},\ref{onLL},\ref{onLR}).  
As before, it is assumed that only particles with largest eigenvalue
of this magnetic field
occupy this state. For the $O(2P)/O(P) \times O(P)$ models at
$\theta=\pi$, there is only one kind of particle (left and right
moving) in the ground state. The Fourier transforms of the kernels are
\begin{eqnarray}
k_1(\omega)&=&1-\widetilde{\phi}_1(\omega) = 
\frac{\sinh((\gamma+1)\pi\Delta\omega)}{\sinh(\gamma\pi\Delta)}
\frac{\cosh((1-2\Delta)\pi\omega/2)}{\cosh(\pi\omega/2)}\\
k_2(\omega)&=&\widetilde{\phi}_2(\omega) = 
\frac{\sinh((\gamma-1)\pi\Delta\omega)}{\sinh(\gamma\pi\Delta)}
\frac{\cosh((1-2\Delta)\pi\omega/2)}{\cosh(\pi\omega/2)}
\end{eqnarray}
where I have again included an extra parameter $\gamma$ to simplify
the analysis. The sigma model kernels are recovered in the limit
$\gamma\to\infty$, giving the exponential factors
$e^{\pi\Delta|\omega|}$ and $e^{-\pi\Delta|\omega|}$ respectively.
Just like the massive case, the $SU(N)/O(N)$ model has two particles
in the ground state, and ends up with the same
kernels.

{}From these explicit forms, one finds remarkably enough 
that $k(\omega)$ in
the massless case is identical to $K(\omega)$ in (\ref{forK}) in the
massive case. The only difference between the equations and
(\ref{forv}) and (\ref{forvpi}) is the extra function
\begin{equation}
\frac{k_1(\omega)}{k_2(\omega)}=
\frac{\sinh((\gamma+1)\Delta\pi\omega)}
{\sinh((\gamma-1)\Delta\pi\omega)}
\label{extra}
\end{equation}
in all three terms.
This indeed means that the energy at $\theta=\pi$ is not the same
as $\theta=0$. However, this extra piece has no effect on
the perturbative contributions, so the energy is still given by
(\ref{power}). This is because the extra piece
introduces no new poles in the integrand in (\ref{forvpi}) (the poles
in (\ref{extra}) are canceled by zeros in $k_-/k_+$). Moreover,
the residues at the poles at $\omega=i(2n+1)/(2\Delta\gamma)$
are the same in the massive and massless cases
(up to an overall sign), because
$$\frac{k_1(i(2n+1)/(2\Delta\gamma))}{k_2(i(2n+1)/(2\Delta\gamma))}=
-1.$$ These residues are what
determine the coefficients $\alpha_n$ in (\ref{power}), so indeed the
perturbative expansions at $\theta=0$ and $\theta=\pi$ are completely
identical. This was established for the sphere sigma model in
\cite{sausage}; here I have extended this proof to two infinite
hierarchies of models.

This completes the identification of the massless $S$ matrices with
the sigma models at $\theta=\pi$. They give the identical perturbative
expansion as the sigma model at $\theta=0$, but the non-perturbative
pieces differ. As an additional check, I have also computed the
$c$-function by computing the free energy with no magnetic field but
at non-zero temperature \cite{meTBA}. This gives the correct
behavior, thus completely confirming this identification.

\section{Flows when $\theta=\pi$}

The main point of this paper is that the stable fixed point in the
sphere sigma model at $\theta=\pi$ is not an isolated instance. There
are at least two infinite hierarchies of models which have this
behavior. Thus current approaches to the problem of disordered
electrons in two dimensions in both the replica approach and the
supersymmetric approach (see \cite{nato} and \cite{zirn} and
references therein) are on sound footing.

One obvious question is such flows happen in other sigma models.  In
this section I will show how to obtain the $Sp(2N)/U(N)$ sigma model
by perturbing the $O(4N)/O(2N)\times O(2N)$ model.  I will use this to
make a conjecture that at $\theta=\pi$ the former has the $Sp(2N)_1$
WZW model as its low-energy fixed point.

First, I will establish a flow between the two integrable hierarchies.
The field $\Phi$ in the $O(2N)/O(N)\times O(N)$ sigma model is an
$2N\times 2N$ traceless real symmetric unitary matrix. It is simple to
see that the field configurations of the $SU(N)/SO(N)$ models are a
subspace of these. The configurations of the former can be written
as
\begin{equation}
\pmatrix{A&B\cr B^T& D}
\label{o2nmat}
\end{equation}
where $A$, $B$, and $D$ are real
$N\times N$ matrices. $A$ and $D$ are symmetric,
and tr$A +$ tr$D=0$. To ensure unitarity the matrices must satisfy
\begin{eqnarray}
\nonumber
A\, A + B^T\, B = D\, D + B^T\, B &=& I\\
\nonumber
A\, B^T + B\, D &=& 0
\end{eqnarray}
where $I$ is the $N\times N$ identity.  The subspace $U(N)/SO(N)$ is
obtained by requiring that $A=-D$ and $B=B^T$.  The
matrix $A+iB$ is indeed a symmetric unitary $N\times N$ matrix, as can be
verified from the preceding matrix relations.
To get the $SU(N)/SO(N)$ sigma model, one must in addition
require det$(A+iB)=1$.
Thus one can flow from the $O(2N)/O(N)\times O(N)$ sigma model to the
$SU(N)/SO(N)$ model adding the potential
\begin{equation}
\lambda \int \left(\hbox{tr}\left[(B-B^T)^2 + (A+D)^2\right]+
|\hbox{det}(A+iB)|^2
\right)
\label{ouflow}
\end{equation}
and making
$\lambda$ large. If $\theta=\pi$ in the former model, then $\theta=\pi$ in
the latter. 
This flow is illustrated in Fig. 1.
\begin{figure}
%{\includegraphics[scale=0.6]{zero.ps}}
\centerline{\epsfxsize=4.0in\epsffile{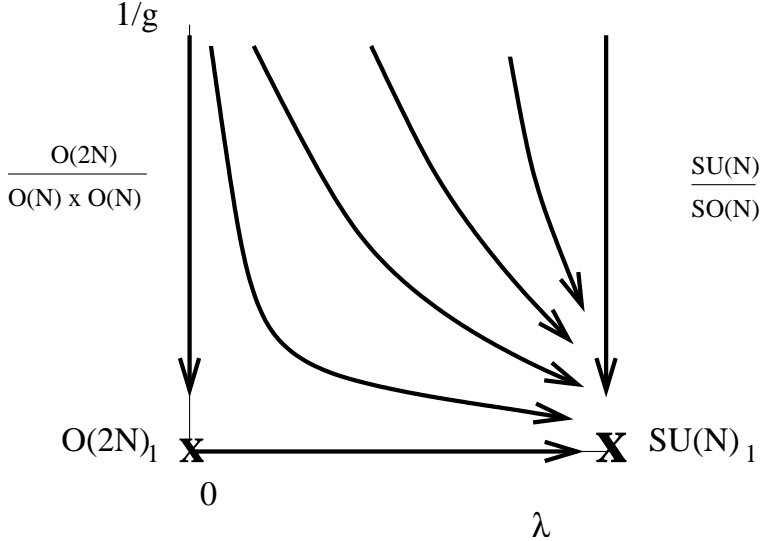}}
\bigskip\bigskip
\caption{The phase diagram of the $O(2N)/O(N)\times O(N)$ sigma model
at $\theta=\pi$
perturbed by coupling $\lambda$ as in (\ref{ouflow}). The $SU(N)_1$
fixed point on the right is at $\lambda=\infty$.  
}
\end{figure}

This flow between the low-energy fixed points can also be seen
explicitly. This is the flow along the $x$-axis in figure 1.  The
$O(2N)_1$ fixed point is equivalent to $2N$ free massless Majorana
(real) fermions. The right-moving fermions are denoted $\psi_R^\alpha$,
while the left movers are denoted $\psi_L^\alpha$, where $\alpha=1\dots 2N$. The
$O(2N)$ symmetry currents are then 
$$j^{\alpha\beta}_R = 2i\, \psi_R^\alpha\psi_R^\beta $$
and likewise for $j^{\alpha\beta}_L$. Because of Fermi statistics,
$j^{\alpha\beta}=-j^{\beta\alpha}$, so there are $N(N-1)/2$ different 
right-moving
currents. (I suppress the $L,R$ subscripts: equations without them are meant
to apply to both $L$ and $R$.)
These currents generate the $O(2N)$ symmetry, which at the
critical point is enhanced to a chiral $O(2N)_L\times O(2N)_R$
symmetry. 
%The Hamiltonian density at the fixed point can be written in terms
%of these currents as
%$H \sum_{i<j} j^{\alpha\beta}_L j^{\alpha\beta}_L + j^{\alpha\beta}_R 
%j^{\alpha\beta}_R$
%where the products are understood to be normal-ordered.
The theory of $2N$ free Majorana fermions is of course equivalent to
a theory of $N$ Dirac fermions $\Psi^a$, defined by
$$\Psi^a =\psi^a + i \psi^{a+N}.$$
for $a=1\dots N$. From the Dirac fermions, one can form the $U(N)$
symmetry currents
$$j^{ab}= \overline{\Psi}^a \Psi^b.$$ 
These $N^2$ right-moving and $N^2$ left-moving
currents generate a
$U(N)_L\times U(N)_R$ subgroup of the full symmetry.

In the terms of WZW models, the equivalence between $N$ Dirac fermions
and $2N$ Majorana fermions is written as
\begin{equation}
O(2N)_1= SU(N)_1 \oplus U(1).
\label{embed1}
\end{equation}
The $U(1)$ symmetry in $U(N)$ is generated by
$$j^0\equiv \sum_{a=1}^N j^{aa}.$$
This can be split off because the generator $j^0$ commutes with all
the generators $j^{ab}$, so the corresponding WZW models
are independent. For $N=2$, this splitting is the simplest
example of spin-charge separation: the $U(1)$ is the charge mode,
while the $SU(2)_1$ are the spin modes.  The
equivalence (\ref{embed1}) is a simple example of what is called a
conformal embedding \cite{GO}. Yet another name often used is
non-abelian bosonization: the free fermions of $O(2N)_1$ are written
in terms of the bosonic fields of the WZW model.  The simplest example of
a conformal embedding is $SU(2)_1=U(1)$. This means that the $SU(2)_1$
can be written in terms of a single boson $\phi$: the $SU(2)$ currents
$j^x,j^y$ and $j^z$ are $\cos\phi,\sin\phi$ and $\partial\phi$
respectively. 

With the conformal embedding/spin-charge separation/non-abelian
bosonization (\ref{embed1}), it is now easy to see how to flow from
$O(2N)_1$ to $SU(N)_1$. One needs to add a perturbation which gives a
gap to the $U(1)$ charge mode but leaves the $SU(N)_1$ untouched. 
If we define the $U(1)$ charge boson $\phi$ by $j^0=\partial\phi$,
the perturbation can
be written as
$$S=S_{O(2N)_1} + \lambda\int \cos(\phi_L+\phi_R),$$ Since the
$SU(N)_1$ currents commute with $j^0$, they are untouched by this
perturbation. Therefore, in the limit $\lambda\to\infty$, the
perturbed theory reaches the $SU(N)_1$ model. Thus the flow between the
two sigma models can be seen in the WZW models as well: essentially all one
needs to do is break the original symmetry appropriately.

This is only one example of a conformal embedding. A complete list
of all such embeddings is given in \cite{embedding}.  
An embedding of particular interest here is
\begin{equation}
SO(4N)_1 = Sp(2N)_1 \oplus SU(2)_N
\label{embed2}
\end{equation}
To realize this embedding
explicitly, it is convenient to write the $SO(4N)$ currents in terms
of real antisymmetric $4N\times 4N$ matrices $T^{\alpha\beta}_{kl}$ where
$${j}^{\alpha\beta}=i\psi^k T^{\alpha\beta}_{kl} \psi^l.$$
Explicitly, $T^{\alpha\beta}_{kl}=\delta^{\alpha}_k\delta^{\beta}_l-
\delta^{\alpha}_l\delta^{\beta}_k$. Then the $SU(2)$ subalgebra is
given by the matrices
\begin{eqnarray}
\nonumber
T^z&=&i\sigma^y\otimes I\\
\nonumber
T^x&=&\sigma^x\otimes Z\\
\nonumber
T^y&=&\sigma^z\otimes Z
\end{eqnarray}
where  $I$ is the $2N\times 2N$ identity, the $\sigma^a$ are the
Pauli matrices, and $Z$ is a fixed real antisymmetric
$2N\times 2N$ matrix which obeys $Z^2=I$. 
Since $T^x$, $T^y$ and $T^z$ are $4N\times 4N$ real
antisymmetric matrices, they are linear combinations of the
$T^{\alpha\beta}$, so they do indeed form an $SU(2)$ subalgebra of
$O(4N)$. The subalgebra which commutes with the $SU(2)$ subalgebra
consists of matrices of the form
\begin{eqnarray}
\nonumber
&&{\cal I}\otimes C\\
\nonumber
&&i\sigma^y\otimes D
\end{eqnarray}
where ${\cal I}$ is the $2\times 2$ identity and $C$ and $D$ are
respectively antisymmetric and symmetric real $2N\times 2N$ matrices.
For these to commute with $T^x,$ $T^y$ and $T^z$, $C$ and $D$ must satisfy
$$[C,Z]=0\qquad \qquad \{D,Z\}=0$$ The subalgebra
consisting of all matrices $C$ and $D$ obeying these constraints
is precisely
$Sp(2N)$: by exponentiating $C$ and $D$ into some matrix $P\equiv e^{C+D}$,
one
finds $P^T\, Z\, P=Z$, which is the defining relation of the group
$Sp(2N)$. Note that because $C$ is real, $P$ is not necessarily unitary.
A unitary matrix can be obtained by exponentiating $iC$ and $D$,
giving the embedding of $Sp(2N)$ in $SU(2N)$.

Therefore, in the embedding (\ref{embed2}),
the currents $J^a=i \psi^k (T^a)_{kl} \psi^l$ ($a=x,y$ or $z$) form the
$SU(2)_N$ current algebra, while the currents 
$\psi^k (I\otimes C)_{kl}\psi^l$
and $\psi^k (i\sigma^y\otimes D)_{kl}\psi^l$ form the $Sp(2N)_1$ 
current algebra.
Thus a flow from the $SO(4N)_1$ fixed point to the $Sp(2N)_1$ fixed
point occurs if one gives a gap to the $SU(2)_N$ modes. This can be
done with the perturbation
$$S=S_{O(4N)_1} + \lambda \int \hbox{tr} \left(J^x_L J^x_R + J^y_L
J^y_R + J^z_L J^z_R\right)$$  
When $\lambda$ goes to $\infty$, the flow reaches the $Sp(2N)_1$
fixed point.  As a tangential comment, note that one can obtain
a fermionic realization of
the $Sp(2N)$ Gross-Neveu model by perturbing $O(4N)_1$
by the $Sp(2N)_1$ currents instead of the $SU(2)_N$ ones.

The question now is if this flow implies a low-energy fixed point in
any sigma model at $\theta=\pi$. My conjecture is that it
does. Namely, consider the $Sp(2N)/U(N)$ sigma model.
Field configurations in this coset space can be realized as a subspace
of the $O(4N)/O(2N)\times O(2N)$ configurations. Matrices in
the latter are of the form (\ref{o2nmat}), where $A$, $B$ and $D$ are
now symmetric
$2N\times 2N$ matrices. To get $Sp(2N)/U(N)$ requires
the restrictions $A=-D$ and $B=B^T$ as before, plus the additional restriction
\begin{equation}
V Z V = Z,
\label{sp}
\end{equation}
where $V\equiv A+iB$. In other words, the $Sp(2N)/U(N)$ subspace
of $O(4N)/O(2N)\times O(2N)$ consists of matrices of the form
$$\frac{1}{2}\pmatrix{V+V^*& i(V-V^*)\cr
i(V-V^*)&-V-V^*}$$
where $V$ is a $2N\times 2N$ symmetric unitary matrix
obeying the condition (\ref{sp}).
One can therefore obtain this model by a perturbation like (\ref{ouflow}).
Here, the perturbation breaks the $O(4N)$ global symmetry down to
$Sp(2N)$. It is giving a large gap to the modes outside the $Sp(2N)$
subgroup, effectively removing them from the theory. This is
just how the flow described above goes from the $SO(4N)_1$
WZW model to the $Sp(2N)_1$ model. It is thus very plausible that the
general flows look like those in Fig.\ 2. 
\begin{figure}
%{\includegraphics[scale=0.6]{zero.ps}}
\centerline{\epsfxsize=4.0in\epsffile{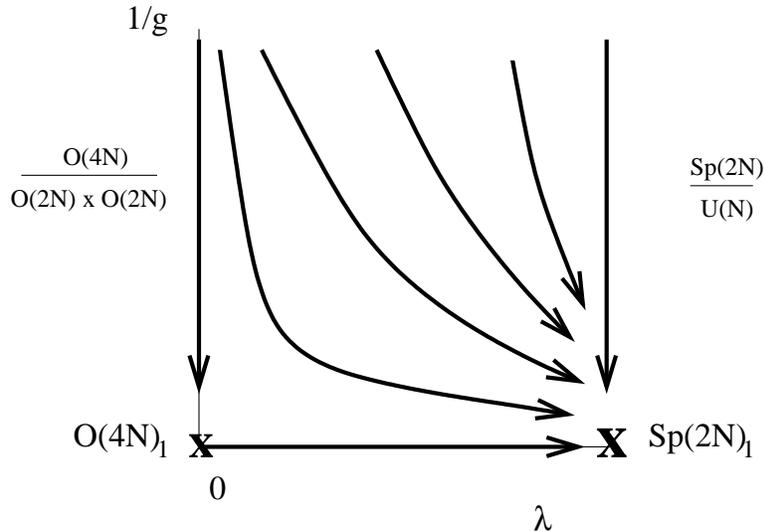}}
\bigskip\bigskip
\caption{The conjectured
phase diagram of the $O(4N)/O(2N)\times O(2N)$ sigma model
at $\theta=\pi$ with symmetry broken to $Sp(2N)$.
}
\end{figure}

This is why I have conjectured that the low-energy fixed
point of the $Sp(2N)/U(N)$ sigma model at $\theta=\pi$
is the $Sp(2N)_1$ WZW model.
This is certainly true at $N=1$, because $Sp(2)=SU(2)$ and this
becomes the flow in the sphere sigma model.
For $N>1$, any non-trivial fixed point in $Sp(2N)/U(N)$
should be a perturbation of the $\theta=\pi$ fixed point of
$O(4N)/O(2N)\times O(2N)$, namely $O(4N)_1$,
which has central charge $2N$.
If there is a non-trivial critical point of $Sp(2N)/U(N)$ when
$\theta=\pi$, it must have central charge less than $2N$, which
leaves only $Sp(2N)_1$

Further evidence for this conjecture comes from the replica
limit $N\to 0$. The resulting model describes the spin quantum
Hall effect \cite{Senthil}, and can also be studied in a supersymmetric
formulation. It is found in \cite{GLR} that
certain correlators are equivalent to those in classical percolation.
{}From this one can extract the density of states exponent
in the disordered model, and indeed it agrees with
the result from taking $N\to 0$ for the corresponding exponent
in $Sp(2N)_1$ \cite{FK} (see also \cite{andre}).
There are many potential subtleties in taking
the replica limit, so this is hardly a proof,
but I take it as a good piece of evidence that the conjecture is true.

\section{Conclusions}

I have shown that several hierarchies of sigma models flow to a
non-trivial low-energy fixed point when $\theta=\pi$. This
provides $SU(N)$ and $O(2N)$-symmetric generalizations of the $SU(2)$
results of \cite{Haldane,ZZtheta}. In another paper, I will
calculate the $c$-functions for these models \cite{meTBA}.

This result is a useful check on the currently-popular approach to
disordered models in two dimensions,
Most recent study has been based on the assumption that
that the picture of \cite{Pruisken} is fairly general, and this paper
shows that the picture holds for the $SU(N)/O(N)$ models (class $C$II)
and the $O(2N)/O(N)\times O(N)$ models (the GSE class). 
All the models discussed in this paper
have WZW models as their non-trivial fixed
points. However, if the model
believed to apply to the integer quantum Hall plateau phase
transition (the $N\to 0$ limit of
$U(2N)/U(N)\times U(N)$ model at $\theta=\pi$ \cite{Pruisken}) has a
fixed point, it does not seem likely that it is of WZW type, as argued
in \cite{Affleck2}. Thus there is still a great deal of interesting
physics yet to be uncovered in sigma models with $\theta=\pi$.

\bigskip\bigskip 

My work is supported by a DOE OJI Award, a Sloan
Foundation Fellowship, and by NSF grant DMR-9802813.

%------------------------- REFERENCES ------------------------------
%\clearpage
%
\renewcommand{\baselinestretch}{1}

\end{document}